\documentclass[aps,prl,reprint,superscriptaddress,amsmath,amssymb,floatfix]{revtex4-2}

\usepackage{graphicx}
\usepackage{dcolumn}
\usepackage{bm}
\usepackage{hyperref}
\usepackage{xcolor, verbatim}

\begin{document}

\title{Seeing through plasmoids: the hidden Petschek structure of fast magnetic reconnection}

\author{Anatoly Spitkovsky}
\email{anatoly@princeton.edu}
\affiliation{Department of Astrophysical Sciences, Princeton University, Princeton, NJ 08544, USA}

\date{\today}

\begin{abstract} 
Whether collisionless magnetic reconnection takes the form of plasmoid chains or
Petschek slow shocks remains unresolved. We show that Petschek's exhaust is the
intrinsic steady state of a reconnecting layer, and plasmoids are a consequence,
not the driver, of fast reconnection. Within a compact skin-depth-scale X-point, finite
Speiser-orbit residence time produces an effectively localized resistivity that
sustains fast reconnection and a macroscopic Petschek outflow. Simultaneously,
Speiser acceleration drives out-of-plane temperature anisotropy, spawning
plasmoids from the X-point via a local Weibel instability rather than due to global
sheet tearing. In pair-plasma kinetic simulations, reducing the X-point anisotropy with
an out-of-plane momentum cap eliminates plasmoids while leaving reconnection
fast and predominantly collisionless. This exposes a purely thermal
exhaust bounded by stationary slow shocks, revealing the Petschek geometry
typically obscured by plasmoids in small-domain simulations. Because the
plasmoid-driving anisotropy subsides beyond a finite distance from the
stagnation point, sufficiently large steady-reconnection systems should
transition from plasmoid-dominated cores to Petschek exhausts across all
magnetizations.
\end{abstract}

\maketitle

\emph{Introduction.---}Magnetic reconnection drives rapid magnetic energy release
across space, solar, and high-energy astrophysical environments, with its rate
intimately linked to the macroscopic layer geometry~\cite{Cassak2017}. In the
Sweet--Parker model~\cite{Parker1957,Sweet1958}, a resistive diffusion region
spanning the macroscopic sheet length $L$ creates a mass-flux bottleneck that
limits the dimensionless reconnection rate to $\beta_{\rm rec} \equiv
v_{{\rm in},0}/v_{A0} \sim S^{-1/2} \ll 1$ at high Lundquist numbers $S$, where
$v_{{\rm in},0}$ and $v_{A0}$ are the upstream inflow and Alfv\'en speeds.
Petschek~\cite{Petschek1964} bypassed this bottleneck by localizing field-line
breaking to a short central diffusion region ($\ell_{\rm diff} \ll L$), while
deflecting and accelerating the bulk inflow across stationary slow-mode shocks
that bound a broadening exhaust. Subsequent resistive magnetohydrodynamic (MHD)
studies showed that uniform resistivity stretches the diffusion region back to
$L$~\cite{Biskamp1986}, whereas anomalous resistivity concentrated near the
X-point sustains Petschek slow shocks robustly~\cite{Ugai1977,Kulsrud2001}.

Collisionless reconnection research took a different historical path. Early
electron--ion simulations~\cite{Shay1998,Birn2001} achieved fast reconnection
($\beta_{\rm rec} \sim 0.1$) and attributed X-point localization to the Hall
effect and dispersive whistler or kinetic Alfv\'en waves.
Subsequent particle-in-cell (PIC) simulations of
electron--positron (pair) plasmas---where the Hall term vanishes by
symmetry---nevertheless remained fast~\cite{Bessho2005,Bessho2007,Hesse2009},
with the off-diagonal pressure tensor supporting an effective collisionless
resistivity. Because those early pair domains were too small to resolve
macroscopic exhausts, it remained unclear what prevents a large-scale kinetic
layer from elongating back into a slow Sweet--Parker sheet. When larger
open-boundary simulations became feasible (both
electron--ion~\cite{Daughton2006} and pair~\cite{Daughton2007,Swisdak2008}),
the thinning sheets were indeed seen to elongate and fragment continually into
secondary magnetic islands (plasmoids). Around the same time, high-$S$
uniform-resistivity MHD sheets were shown to bypass the Sweet--Parker
bottleneck via the plasmoid (secondary tearing)
instability~\cite{Loureiro2007,Bhattacharjee2009,Uzdensky2010}, which breaks a
long layer into shorter, faster-reconnecting segments. Together with large
relativistic PIC simulations that were always dominated by continual plasmoid nucleation,
mergers, and secondary X-points~[e.g.,~\onlinecite{Sironi2014,Sironi2016,OrtunoMacias2020}],
these results established the paradigm that collisionless pair reconnection
requires plasmoid chains to keep diffusion regions short and remain fast,
without forming a stationary Petschek exhaust. Recently, however, very
large open-domain pair simulations in the subrelativistic regime~\cite{Robbins2025} revealed that on
macroscopic scales
reconnection relaxes into a broadening Petschek-like exhaust bounded by slow
shocks despite persistent central plasmoids, leaving open whether fast
collisionless reconnection is driven by plasmoid chains or Petschek's mechanism
and how the two are related. 

In this Letter, we reconcile these results by showing that steady 2D
antiparallel collisionless pair reconnection is intrinsically a Petschek process
driven by effectively localized Speiser-orbit
resistivity~\cite{Speiser1965,Speiser1970} at a compact primary stagnation point,
whereas plasmoids are a consequence rather than the cause of fast
reconnection. Building on recent demonstrations of Weibel-driven plasmoid
birth~\cite{Goodbred2026} and large-scale Petschek exhausts~\cite{Robbins2025},
we use a Speiser residence-time argument to analytically derive the distribution function of
accelerating charges in an active X-point and show that the reconnection electric field
drives temperature anisotropy that triggers the Weibel instability.  
We demonstrate this causality directly with 2D PIC simulations by showing 
that artificial reduction of anisotropy near the stagnation point 
completely eliminates plasmoids while leaving reconnection fast and 
predominantly pressure-tensor supported, directly uncovering the underlying
thermal Petschek exhaust bounded by slow shocks that is typically hidden 
underneath the vigorous plasmoid dynamics in small PIC simulations. 

\emph{Speiser residence time, localized resistivity, and Weibel anisotropy.---}Consider
steady 2D antiparallel reconnection in a pair plasma without a guide field after
initial transients have relaxed, establishing a clear inflow along $y$,
bipolar outflow along $x$, reversing magnetic field $\pm B_0 \hat{\mathbf{x}}$,
and current along $z$. In any open layer sustaining a global bipolar outflow,
there exists at least one primary flow-dividing stagnation region. For upstream
pair density $n_0 = n_{+0} + n_{-0}$, plasma frequency $\omega_{p0} \equiv
(4\pi n_0 e^2 / m_e)^{1/2}$, skin depth $d_e \equiv c/\omega_{p0}$, and
magnetization $\sigma_0 \equiv B_0^2/(4\pi n_0 m_e c^2)$, the upstream Alfv\'en speed
is $v_{A0} = c[\sigma_0/(1+\sigma_0)]^{1/2}$. In steady reconnection, a uniform
reconnection electric field $E_z$ permeates the domain. Within the unmagnetized
X-point diffusion region of half-thickness $\delta$ and half-length $L_x$,
particles executing Speiser meandering oscillations across
$y=0$~\cite{Speiser1965} accelerate along $\pm \hat{\mathbf{z}}$ as
$dp_z/dt = q E_z$. Because the accelerating outflow $v_x(x) \approx v_{\rm out}(x/L_x)$
sweeps particles out of the X-region at a constant rate
$\partial_x v_x \approx v_{\rm out}/L_x \equiv \tau_{\rm res}^{-1}$, particle
ages $\tau$ follow a normalized exponential distribution
$P(\tau) = \tau_{\rm res}^{-1} e^{-\tau/\tau_{\rm res}}$ ($\tau \ge 0$; \hyperref[app:residence]{Appendix~A}).
Positrons ($+$) and electrons ($-$) thus acquire mean out-of-plane drift
momenta $\langle p_z \rangle_\pm = \pm p_d$, where
$p_d \equiv \int_0^\infty e E_z \tau P(\tau)\,d\tau = e E_z \tau_{\rm res} = \bar{\gamma} m_e v_d$~\cite{Speiser1970}, with mean drift speed $v_d \equiv |\langle v_z \rangle_\pm|$ and velocity-weighted carrier Lorentz factor $\bar{\gamma} \equiv \langle \gamma v_z \rangle_\pm / \langle v_z \rangle_\pm$.

Combining the current density $J_z = e n v_d$ with Amp\`ere's law
$J_z \sim c B_{x,{\rm in}} / (4\pi \delta)$ (where $B_{x,{\rm in}} \lesssim B_0$ is the local reconnecting field at the inflow edge of the diffusion region) yields the non-ideal electric field at the X-point~\cite{Speiser1970,Lyons1985,Selvi2023}:
\begin{equation}
E_z \sim \frac{4\pi d_{e,{\rm rel}}^2}{c^2 \tau_{\rm res}} J_z \sim \frac{v_{\rm out} B_{x,{\rm in}}}{c} \left(\frac{d_{e,{\rm rel}}}{\delta}\right)\left(\frac{d_{e,{\rm rel}}}{L_x}\right),
\label{eq:speiser_Ez}
\end{equation}
where $d_{e,{\rm rel}} = \sqrt{\bar{\gamma}}\,d_e$ is the local relativistic skin
depth. 
Physically, particles decouple from the field lines in the diffusion region and
accelerate along the reconnection electric field, but are
continuously advected into the outflow before the field can be fully shorted out. 
This slippage due 
to finite residence time near the X-point keeps the electric
field non-zero and functions as effective resistivity. 
In fluid moment language, the transport of out-of-plane momentum appears as the
divergence of the off-diagonal pressure tensor,
$E_z = (\nabla \cdot \mathbf{P})_z / (en)$~\cite{Vasyliunas1975,Kuznetsova1998,Hesse1999,Hesse2009}.
While it is tempting to conclude that the pressure tensor causes the non-ideal electric field,
in fact, these pressure moments arise simply as a consequence of collisionless 
particle dynamics in the accelerating field of the layer. 

Matching $E_z = (v_{\rm in}/c)B_{x,{\rm in}}$ (where $v_{\rm in}$ is the local inflow-edge speed) with continuity $v_{\rm in}/v_{\rm out} \sim
\delta/L_x$ in Eq.~(\ref{eq:speiser_Ez}) cancels $L_x$ and fixes
$\delta \sim d_{e,{\rm rel}}$, so Amp\`ere's law gives $\bar{\gamma}(v_d/c)^2 \sim \sigma_{\rm in}$ (yielding $v_d \sim v_{A,{\rm in}} < v_{A0}$ and
$\bar{\gamma} \sim 1 + \sigma_{\rm in}$, in terms of the inflow-edge magnetization $\sigma_{\rm in}$ and Alfv\'en speed $v_{A,{\rm in}}$), leaving the aspect ratio $\delta/L_x$ to macroscopic force balance.
Because the effective Speiser resistivity ($\propto v_{\rm out}/L_x$) grows as $L_x$ shrinks, localized X-point dissipation imposes no resistive bottleneck against opening the layer; however, bending the upstream field lines more sharply increases upward magnetic tension, which weakens $B_{x,{\rm in}} < B_0$ (via a fast-mode rarefaction as the inflow accelerates toward the X-point) and reduces the local Alfv\'en and slingshot outflow speeds.
Together with exhaust backpressure opposing the outflow, this trade-off limits the global reconnection rate to $\beta_{\rm rec} \approx 0.1$~\cite{Petschek1964,Hesse2009,Liu2017,Cassak2017,Liu2020,Liu2022,Goodbred2022,Selvi2023}. Because thinning the layer to the skin-depth scale
$\delta \sim d_{e,{\rm rel}}$ drives $v_d \sim v_{A,{\rm in}} \lesssim v_{A0}$, particle
demagnetization coincides with the current reaching the Alfv\'enic scale
$J_A \equiv e n v_{A0}$, with
$|J_z|/J_A \sim (n_0/n)(d_{e,{\rm rel}}/\delta)$ by Amp\`ere's law. There, with
$\tau_{\rm res} \sim d_{e,{\rm rel}}/(\beta_{\rm rec} v_{A0})$,
Eq.~(\ref{eq:speiser_Ez}) gives an effective resistivity
$\eta_{\rm eff} \equiv E_z^*/J_z \sim (4\pi/c^2)\,\beta_{\rm rec} v_{A0} d_{e,{\rm rel}} = \beta_{\rm rec} B_0/(e n_0 c)$
(magnetic diffusivity $D_m \sim \beta_{\rm rec} v_{A0} d_{e,{\rm rel}}$), where
$E_z^* \equiv E_z + (\mathbf{v}\times\mathbf{B})_z/c$ is the non-ideal
electric field. Because X-point pressure depletion and force balance keep the unmagnetized
diffusion region compact ($L_x \ll L$)~\cite{Liu2020,Goodbred2022},
remagnetization in the normal field $B_y$~\cite{Selvi2023} restores ideal
frozen-in motion ($E_z^* \to 0$) beyond $|x| \gtrsim L_x$, while exhaust
broadening and mass loading dilute $|J_z| \ll J_A$. Thus $\eta_{\rm eff}$ is
effectively localized to $|x| \lesssim L_x \ll L$; viewed globally, this forces
the large-scale flow ($|x| \gg L_x$) into a Petschek exhaust bounded by
slow-mode shocks, just as in localized-resistivity
MHD~\cite{Ugai1977,Biskamp1986,Kulsrud2001}.

\begin{figure*}[t]
\centering
\includegraphics[width=0.95\textwidth]{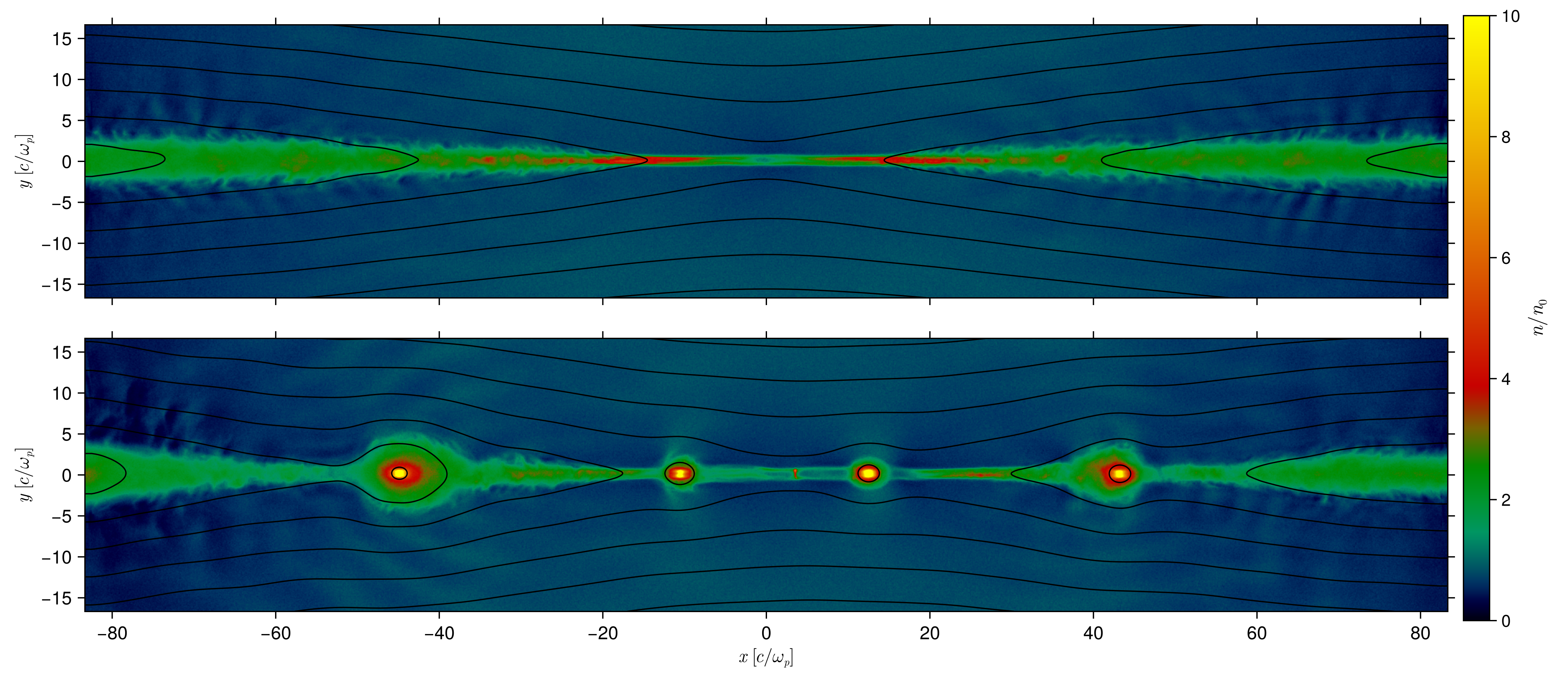}
\caption{Normalized density $n/n_0$ (color) and in-plane magnetic field lines
(black contours) in 2D pair-plasma reconnection ($\sigma_0 = 0.1$). Top:
Capping $|p_z| \le 2.3 p_d$ in a thin strip across the X-point reduces anisotropy,
eliminating plasmoids and revealing a Petschek exhaust bounded by slow shocks.
Bottom: Uncapped run, where Weibel-generated plasmoids born at the X-point
(note the nascent sub-skin-depth plasmoid near $x \approx 3.5\,d_e$) dominate
the exhaust.} 
\label{fig:morphology}
\end{figure*}

Simultaneously, Speiser acceleration drives strong momentum anisotropy in the
stagnation zone. Near $x \to 0$, inflowing particles have not yet accelerated
along $x$, keeping their in-plane spread at the upstream thermal level
$\langle p_x^2 \rangle = \langle p_{x0}^2 \rangle \approx m_e T_0$. Along $z$,
combining $p_z(\tau) = q E_z \tau = \pm p_d (\tau/\tau_{\rm res})$ with $P(\tau)$
does \emph{not} yield narrow beams at $\pm p_d$: because newly arrived particles
($\tau \to 0$) are most numerous, each species forms a one-sided exponential
$f_\pm(p_z) = (n_0 / 2 p_d) e^{\mp p_z/p_d}$ for $\operatorname{sgn}(p_z) = \pm 1$
(see \hyperref[app:laplace]{Appendix~B}).
Summing both species gives a Laplace distribution,
\begin{equation}
F(p_z) \equiv f_+(p_z) + f_-(p_z) = \frac{n_0}{2 p_d} e^{-|p_z|/p_d},
\label{eq:laplace_dist}
\end{equation}
peaking at $p_z = 0$ with variance $\langle p_z^2 \rangle = 2 p_d^2 \gg \langle p_{x0}^2 \rangle$ (while single-particle orbit treatments give slightly different far-tail shapes~\cite{Bulanov1975,Zenitani2001,Larrabee2003,Bessho2012}, Eq.~(\ref{eq:laplace_dist}) accurately captures the core and bulk moments).

Plasma with such anisotropy is unstable to the Weibel~\cite{Weibel1959} filamentation mode
with wavevector $\mathbf{k} = k_x\hat{\mathbf{x}}$ along the sheet (whose
$\delta B_y$ perturbations pinch the layer into filaments). Defining
dimensionless relativistic temperature components $\Theta_i \equiv \langle u_i^2/\gamma \rangle$
[with $\mathbf{u} = \mathbf{p}/(m_e c)$ and $u_d \equiv p_d/(m_e c) \sim \sqrt{\sigma_0(1+\sigma_0)}$]
and effective skin depth $d_{e,{\rm rel}}^{-2} = (\omega_{p0}^2/c^2)\langle 1/\gamma \rangle$
(agreeing with $\sqrt{\bar{\gamma}}\,d_e$ up to a weak $\sqrt{\ln\sigma_0}$ factor at
$\sigma_0 \gg 1$), the relativistic Vlasov marginal-stability wavenumber $k_c$
satisfies $k_c^2 d_{e,{\rm rel}}^2 \simeq A$~\cite{Goodbred2026}, where
\begin{equation}
A \equiv \frac{\Theta_z}{\Theta_x} - 1 = \frac{\langle u_z^2/\gamma \rangle}{\langle u_{x0}^2 \rangle \langle 1/\gamma \rangle} - 1.
\label{eq:anisotropy}
\end{equation}
Across all $\sigma_0$, $\Theta_z$ is dominated by older tail particles
($\tau \gtrsim \tau_{\rm res}$, $|p_z| \gtrsim p_d$). For $\sigma_0 \ll 1$ ($\gamma \approx 1$),
Eq.~(\ref{eq:laplace_dist}) yields $\Theta_z \approx 2 u_d^2$ (with the integrand
$p_z^2 F(p_z)$ peaking at $|p_z| = 2 p_d$, $\tau = 2\tau_{\rm res}$) and
$\langle 1/\gamma \rangle \approx 1$, giving
$A = 2 p_d^2 / \langle p_{x0}^2 \rangle - 1 \sim 4/\beta_{p0} \gg 1$ (with
$\beta_{p0} \equiv 8\pi n_0 T_0 / B_0^2$). For a cold upstream at
$\sigma_0 \gg 1$ ($T_0 \ll m_e c^2$, $u_d \sim \sigma_0 \gg 1$),
$\gamma(\tau) \approx |u_z(\tau)|$ yields $\Theta_z \simeq u_d$ from the
$\tau \sim \tau_{\rm res}$ tail, whereas young particles ($\tau \ll \tau_{\rm res}$,
$\gamma \sim 1$) weight $\langle 1/\gamma \rangle \simeq \ln(2u_d)/u_d$, giving
$A \simeq u_d^2 / [\langle u_{x0}^2 \rangle \ln(2u_d)] \sim \sigma_0 / (\beta_{p0} \ln\sigma_0) \gg 1$.

Unlike studies of imposed anisotropy in static initial sheets~\cite{Coppi1983,Chen1984,Karimabadi2004,Quest2010}
or downstream exhaust and jet-front filamentation~\cite{Swisdak2008,Zenitani2008,Liu2009}, here $A \gg 1$ is
continuously generated by Speiser acceleration at the active X-point.
Moreover, in addition to $p_z$ stretching, Speiser phase-space focusing at the
stagnation point correlates $p_x$ with $p_z$, narrowing the $p_x$ spread of the
highest-$|p_z|$ particles. This further enhances the effective anisotropy right
at the X-point, increasing the cutoff wavenumber $k_c$ and boosting Weibel
growth on skin-depth and sub-skin-depth scales ($\lambda_x \lesssim d_{e,{\rm rel}}$), 
so that modes outgrow local outflow stretching ($\partial_x v_x \sim \tau_{\rm res}^{-1}$)~\cite{Bulanov1978}
and pinch into plasmoids~\cite{Goodbred2026}. Moving away from the X-point,
remagnetization in $B_y$, mass accumulation, and pickup heating rapidly broaden
$p_x$ and reduce $\Theta_z/\Theta_x$, confining plasmoid nucleation to the
vicinity of the X-point and shutting off the Weibel instability beyond the
isotropization distance
$\ell_{\rm iso} \sim (1+\sigma_0) d_e / \beta_{\rm rec}$~\cite{Goodbred2026}.

Because anisotropy is carried by high-$|p_z|$ particles that periodically get trapped and advected
by plasmoids, refilling of the tail over $\Delta p_z \sim p_d$ sets the
plasmoid production timescale at $\tau_{\rm res} = p_d/(e E_z) \sim [(1+\sigma_0)/(\beta_{\rm rec}\sqrt{\sigma_0})]\,\omega_{p0}^{-1}$.
Reducing the high-$|p_z|$ tail near the X-point should therefore
lower the Weibel anisotropy without disrupting the bulk sheet current and 
suppress plasmoid formation while leaving fast reconnection
intact.

\begin{figure*}[t]
\centering
\includegraphics[width=0.99\textwidth]{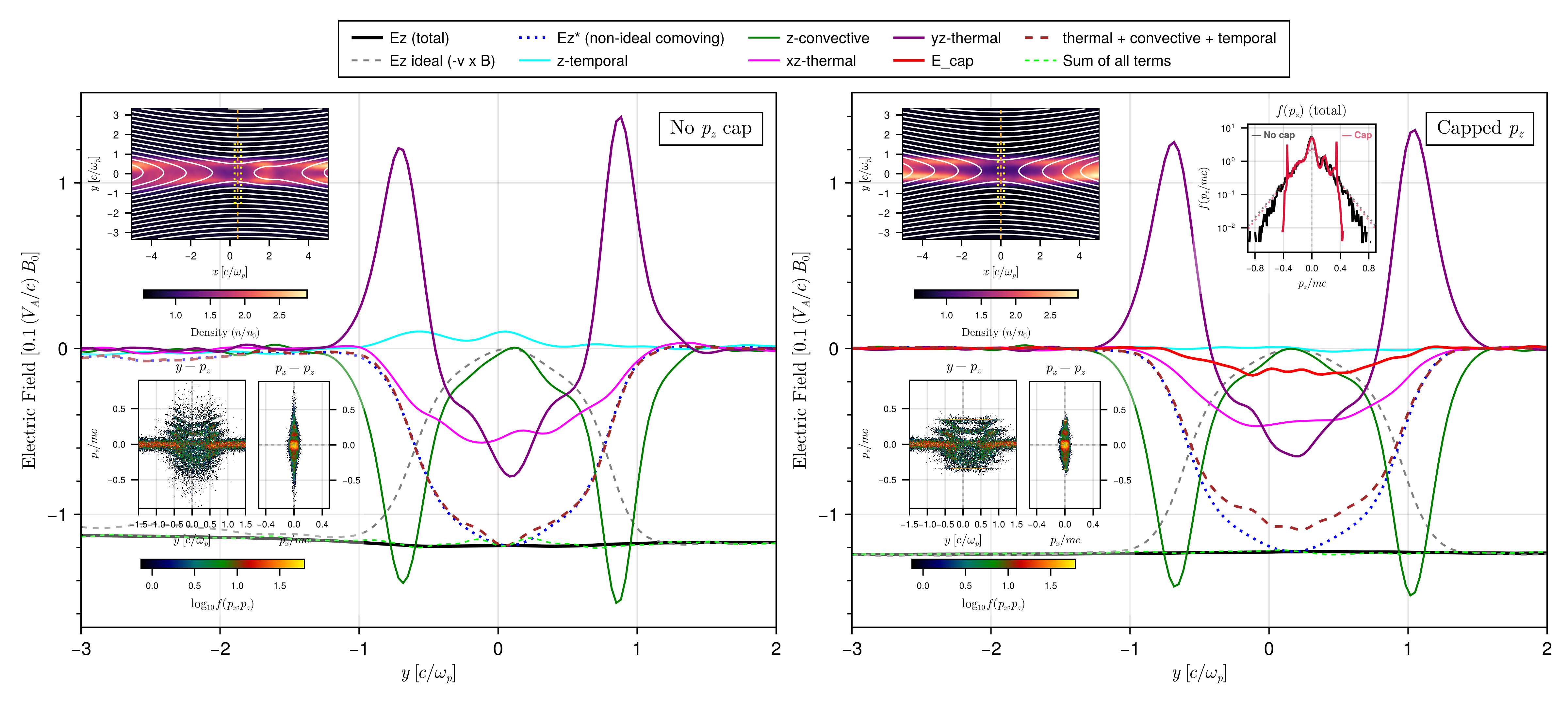}
\caption{Transverse profiles of the generalized Ohm's law
[Eq.~(\ref{eq:ohms_law})] across the X-point without capping (left) and with a
$|p_z| \le 2.3 p_d$ cap in a thin strip of width $2w = 0.33\,d_e$ across the X-point (right), normalized to
$0.1\,(v_{A0}/c)B_0$. Insets show near-X-point density and field lines (with orange
line showing the cut for $E_z$) 
and $y\text{--}p_z$ and $p_x\text{--}p_z$ phase spaces
(shared colorbar) extracted from the yellow box on the density plot. 
With the cap enabled, the off-diagonal
pressure-tensor terms still support $90\%$ of the integrated $E_z^*$, while
$E_{\rm cap}$ (red) contributes only $10\%$. Distribution functions $f(p_z)$ from the yellow box region are shown in the
top-right inset alongside an exponential fit (dotted). }
\label{fig:ohms_law}
\end{figure*}

\emph{Numerical setup.---}We test this prediction using 2D fully kinetic
TRISTAN-MP~\cite{Spitkovsky2005} PIC simulations of an electron--positron
plasma with $\sigma_0 = 0.1$, upstream plasma beta $\beta_{p0} = 0.04$, and
zero guide field.
The domain spans $N_x \times N_y = 5000 \times 3840$ cells with $d_e = 30$ cells
($167\,d_e \times 128\,d_e$) and 16 particles per cell per species,
using absorbing outflow boundaries along $x$ and inflow boundaries along $y$.
Reconnection is initiated at $(x,y)=(0,0)$ by a localized pressure depletion in
a Harris sheet equilibrium of initial thickness $3 d_e$ and overdensity of $4 n_0$. 
The simulation runs for $\sim 3500\,\omega_{p0}^{-1}$ to establish the steady state. 
Upstream boundaries adjust the inflow velocity to maintain constant density far 
upstream and inject new magnetic flux. 

To regulate stagnation-zone anisotropy without disrupting the bulk current
carriers, we cap out-of-plane particle momenta to
$|p_z| \le p_{\rm cap} \equiv \xi p_d$ inside a narrow sub-skin-depth strip
$|x| \le w$ ($w = 0.167 d_e$, spanning all $y$) centered on $x = 0$,
while leaving $p_x$ and $p_y$ untouched. Confining the cap to this thin X-point
strip is essential: capping $|p_z|$ everywhere in the domain clips the hot
thermal tail of the downstream exhaust and inhibits outflow expansion,
whereas the strip targets only the unmagnetized Speiser tail while leaving the
exhaust completely free. Imposing a momentum ceiling is equivalent to introducing an effective
electric field $E_{\rm cap}$ that acts only on the fastest tail particles
($|p_z| = p_{\rm cap}$) to cancel their acceleration by $E_z$, thereby
modifying the moments of $F(p_z)$ (\hyperref[app:cap]{Appendix~C}). For our fiducial cap $p_{\rm cap} = 2.3 p_d$ ($\xi = 2.3$),
$90\%$ of the main current moment $\langle |p_z|\rangle$ is retained, and 
$E_{\rm cap}/E_z = e^{-2.3} \approx 10\%$, while 
$\langle p_z^2\rangle$ drops by $33\%$. Because the
clipped $|p_z| > 2.3 p_d$ tail contains the coldest-$p_x$, Speiser-focused
particles, the cap suppresses the effective anisotropy that drives short-wavelength modes 
far more strongly---enough to push Weibel growth below the local
outflow-stretching threshold.

\emph{Plasmoid suppression and the underlying Petschek state.---}Figure~\ref{fig:morphology}
contrasts the baseline uncapped run (bottom) with the capped run (top). Without
the momentum cap, strong X-point anisotropy continually spawns plasmoids that
nucleate on sub-skin-depth scales near the X-point (visible as the nascent
vertical density strip near $x \approx 3.5\,d_e$ in the bottom panel of
Fig.~\ref{fig:morphology}). Plasmoids then grow as they advect into the exhaust, and can
displace the primary X-point. With the thin-strip momentum cap enabled,
trimming the Speiser-focused tail completely eliminates plasmoid nucleation
and unambiguously reveals a stationary, wedge-shaped Petschek exhaust bounded
by slow-mode shocks. As $p_{\rm cap}$ is raised toward $\approx 3 p_d$,
reconnection remains fast at $\beta_{\rm rec} \approx 0.11\text{--}0.12$ until
anisotropy crosses the effective Weibel threshold and plasmoids reappear at
the X-point.
 
\emph{Non-ideal electric field budget.---}To verify that momentum capping does
not merely replace collisionless reconnection with localized resistive-MHD
reconnection~\cite{Ugai1977,Biskamp1986}, we evaluate the 
out-of-plane generalized Ohm's law 
across the X-point:
\begin{equation}
\begin{aligned}
E_z + \frac{1}{c}(\mathbf{v} \times \mathbf{B})_z &= \frac{1}{e n}\sum_{s=\pm} s\,\Bigl[(\nabla \cdot \mathbf{P}_s)_z \\
&\quad + n_s(\partial_t + \mathbf{v}_s \cdot \nabla)\bar{p}_{s,z}\Bigr] + E_{\rm cap},
\end{aligned}
\label{eq:ohms_law} 
\end{equation}
where $s = \pm 1$ labels positrons and electrons, $n = n_+ + n_-$,
$\mathbf{v} = (n_+\mathbf{v}_+ + n_-\mathbf{v}_-)/n$, $\bar{p}_{s,z} \equiv
\langle p_z\rangle_s$ is each species' fluid mean momentum, and $E_{\rm cap}$
is the effective electric field exerted by the momentum cap (evaluated from the
momentum removed in the simulation). 
Figure~\ref{fig:ohms_law} plots vertical cuts of Eq.~(\ref{eq:ohms_law}) through
the X-point for the uncapped (left) and capped (right) runs. In both, the total
reconnection field $E_z$ (solid black) is nearly uniform at
$\beta_{\rm rec} \approx 0.11\text{--}0.12$. For $|y| \gtrsim d_e$, $E_z$ is
balanced by the ideal motional field $-(\mathbf{v}\times\mathbf{B})_z/c$ (dashed
gray), while inside $|y| \lesssim d_e$ the non-ideal field
$E_z^* = E_z + (\mathbf{v}\times\mathbf{B})_z/c$ (dotted blue) is
governed by kinetic momentum transport. Across $0.4 \lesssim |y|/d_e \lesssim 1$,
inward convective inertia $\sum_s s(n_s/n)(\mathbf{v}_s\cdot\nabla\bar{p}_{s,z})/e$
(green) opposes $\sum_s s\,\partial_y P_{s,yz}/(en)$ (purple) as particles
enter Speiser orbits, whereas in the core ($|y| \lesssim 0.5\,d_e$) both
off-diagonal pressure-tensor derivatives $\sum_s s\,\partial_y P_{s,yz}/(en)$
and $\sum_s s\,\partial_x P_{s,xz}/(en)$ (magenta) support $E_z^*$. With the
$|p_z| \le 2.3 p_d$ cap enabled (right),
the off-diagonal pressure-tensor divergence
$\sum_s s(\nabla\cdot\mathbf{P}_s)_z/(en)$ still supplies $90\%$ of the
$y$-integrated $E_z^*$ (and $\sim 90\%$ right at $y = 0$), while $E_{\rm cap}$
(red) is smoothly distributed across the meandering layer ($|y| \lesssim d_e$)
with a small amplitude ($\sim 10\%$ at $y=0$) and an integrated
contribution of only $10\%$, matching Eq.~(\ref{eq:cap_moments}) 
in \hyperref[app:cap]{Appendix~C} ($e^{-2.3} = 10.0\%$). 
The
phase-space and $f(p_z)$ insets confirm that the
unimodal Laplace core [Eq.~(\ref{eq:laplace_dist})] and bulk sheet current
remain intact while the cap selectively trims the tapered, cold-$p_x$
high-$|p_z|$ needle of Speiser-focused particles (visible in the uncapped
$p_x\text{--}p_z$ inset), piling them up at $|p_z| = p_{\rm cap}$. Thus fast
reconnection remains collisionless and pressure-tensor supported while the
secondary Weibel instability is quenched.

\emph{Flow geometry.---}In this laminar steady state, the non-ideal X-point ($E_z^* \neq 0$) is compact ($\delta \approx d_e$, $L_x \approx 3\,d_e$) with a wide local separatrix slope ($\sim 0.3$), while upstream field weakening ($B_{x,{\rm in}} \approx 0.56\,B_0$) and inner-exhaust backpressure limit the global rate to $\beta_{\rm rec} \approx 0.12$. Further downstream, transverse cuts across the stationary broadening exhaust exhibit a sharp density jump anti-correlated with a drop in $|B_x|$ at nearly uniform normal field $B_y \approx 0.1 B_0$. While in
our fiducial domain ($|x| \lesssim 80\,d_e$) the exhaust thickness remains
comparable to the kinetic slow-shock transition width (set by the downstream
gyroradius in $B_y$, $\rho_{e,y} \sim 5\,d_e$), a larger capped simulation (not
shown) resolves the separated shock structure at $|x| \approx 400\,d_e$
downstream of the X-point, displaying bifurcated boundary $J_z$ layers, the
characteristic switch-off slow-shock density compression ratio
$n_d/n_{\rm up} \approx 2.5$, and partial-switch-off shelves
($|B_x| \approx 0.45 B_0$) arising from residual exhaust pressure
anisotropy~\cite{Liu2012,Higashimori2012,Innocenti2015}, thus achieving 
the same Petschek state obtained on $>1000 d_e$ scales in plasmoid-producing simulations by Robbins \&
Spitkovsky~\cite{Robbins2025}.

\emph{Discussion.---}Steady collisionless reconnection is inherently fast
without requiring plasmoids: when current sheets thin to skin-depth scales and
pressure depletion localizes the unmagnetized X-point
($L_x \ll L$)~\cite{Liu2020,Goodbred2022}, finite Speiser residence time
provides an effectively localized resistivity ($E_z^* \to 0$ outside
$|x| \lesssim L_x$) that breaks line freezing and, on macroscopic scales,
forces the large-scale flow into a Petschek exhaust bounded by slow shocks.
By focusing on the steady state with one main stagnation region, our 
argument bypasses the onset problem
involving global tearing of initially thick layers or external compression.
Plasmoid birth near the primary X-point is driven by local temperature
anisotropy continually regenerated by the collisionless dynamics of Speiser orbits. This reflects the
Weibel limit of anisotropic collisionless tearing~\cite{Schoeffler2026}, rather
than global MHD-like tearing of an elongated Sweet--Parker sheet. In fact, 
though they look very similar and ultimately grow by attraction of like currents,  
plasmoids seen in kinetic simulations may be a different phenomenon from the 
tearing-generated plasmoids observed in MHD simulations, or at least they are seeded
by very different physical processes (local kinetic anisotropy vs. global sheet tearing).  

In PIC simulations, a collisionless X-point undergoing active reconnection always 
tries to relax to the natural steady state of 
a broadening Petschek exhaust, but this process is continuously disrupted by the 
formation of plasmoids as the reconnection electric field drives plasma anisotropy in the 
diffusion region. 
Advected plasmoids act as an ``anisotropy release valve'' that periodically vents
the high-$|p_z|$ tail and resets the X-point. Plasmoids violate strict stationarity and complicate the evolution 
on small scales through mergers and X-point recoil.  On large scales ($L \gg \ell_{\rm iso}$), however, since Weibel plasmoid birth ceases
beyond $\ell_{\rm iso}$~\cite{Goodbred2026}, a steady layer with one primary
stagnation region should transition to a thick Petschek exhaust with plasmoids
embedded inside or gently modulating it. This explains why
smaller boxes ($L \lesssim \ell_{\rm iso}$) appear purely plasmoid-dominated, 
while large subrelativistic domains recover Petschek geometry~\cite{Robbins2025}. We
expect a similar transition for $L \gg \ell_{\rm iso}$ to happen in relativistic
($\sigma_0 > 1$) layers, to be tested in future work.
Strong guide fields ($B_z \sim B_0$) resist plasmoid formation by
magnetizing current carriers and suppressing lateral clumping needed for Weibel
filamentation~\cite{Karimabadi2005,Liu2020}. In three dimensions, drift-kink and
oblique tearing modes introduce flux-rope kinking and turbulence in antiparallel
layers~\cite{Daughton2011,Sironi2014}, though localized X-line dissipation and
downstream isotropization should still favor broad exhausts on large scales,
especially with modest guide fields. Only a broadening exhaust enables the
current sheet to process the mass flux that is increasing
with distance from the X-point without requiring enormous plasma compression.

Although our simulations use pair plasma, similar physics should apply to
electron--ion reconnection: ions decouple earlier, but field lines remain tied
to electrons down to electron skin-depth scales, where electron 
anisotropies should similarly seed plasmoids near the X-point~\cite{Daughton2006}, while effectively localized
electron Speiser resistivity breaks field lines at the X-point and Hall physics bridges the
electron and ion scales to establish a fast, mass-ratio-independent rate and
Petschek geometry on scales well above the ion skin depth. 

Downstream particle spectra in our plasmoid-free run are purely
thermal, confirming that plasmoid dynamics---rather than slow shocks---drives
nonthermal acceleration in low-$\sigma_0$ reconnection~\cite{Robbins2025}, since 
only a fraction $\sim L_x/L \ll 1$ of inflowing plasma 
crosses the X-point while the
bulk is heated at slow shocks. 
If active plasmoids in relativistic reconnection
are likewise confined within $\sim \ell_{\rm iso}$ of primary X-lines,
astrophysical reconnection may face geometric constraints on nonthermal
acceleration efficiency, particularly in steady-reconnection scenarios.
In astrophysics, strong Compton cooling can likewise suppress plasmoids and
particle injection when radiation reaction caps the Speiser tail at
$\gamma_{\rm rad} \lesssim \text{few}\,\sigma_0$ (cf.~\cite{Sironi2020}). In
laboratory experiments, reducing X-point anisotropy via collisional or
laser-induced electron drag, or with larger parallel plasma $\beta$, can have the
same effect. 

Finally, the Speiser residence-time picture clarifies how to parameterize
kinetic effective resistivity in macroscopic MHD simulations across arbitrary
magnetizations. Empirical relativistic prescriptions~\cite{Moran2025,Ripperda2026}
activate resistivity at the charge-starvation threshold
$J_{\rm thresh} = e n c$ ($v_d \to c$), which does not extend to
$\sigma_0 \lesssim 1$, where Amp\`ere's law at $\delta \sim d_{e,{\rm rel}}$
limits carrier drift to $v_d \lesssim v_{A0} \ll c$. Because the X-point Speiser
resistivity $\eta_{\rm eff} \sim \beta_{\rm rec} B_0/(e n c)$ has the same form
at all $\sigma_0$ (giving $E_z = \eta_{\rm eff} J_A \sim \beta_{\rm rec}(v_{A0}/c)B_0$),
triggering it at the Alfv\'enic threshold $J_A = e n v_{A0}$ instead of $e n c$
generalizes these prescriptions to arbitrary magnetizations, activating fast
reconnection when the sheet thins to the effective skin-depth scale and shutting it
off downstream as mass loading drives $|\mathbf{J}|/J_A \ll 1$.

\begin{acknowledgments}
We acknowledge invaluable discussions with Matthew Goodbred, Adam Robbins, 
Alexander Chernoglazov, Alexander Philippov, Dmitri Uzdensky, 
Ehud Nakar, Matthew Kunz, Lorenzo Sironi, Amitava Bhattacharjee, and Ellen Zweibel.
This work is dedicated to the memory
of Nuno Loureiro, who inspired and encouraged this research.
Simulations were performed using computational resources at Princeton
University. This research is supported by NSF (grant PHY-2206607, MPPC)
and Simons Foundation (grant MP-SCMPS-0000147).
Google Gemini was used for editing text and analysis scripts.
\end{acknowledgments}

\bibliography{references}

\begin{thebibliography}{56}%
\makeatletter
\providecommand \@ifxundefined [1]{%
 \@ifx{#1\undefined}
}%
\providecommand \@ifnum [1]{%
 \ifnum #1\expandafter \@firstoftwo
 \else \expandafter \@secondoftwo
 \fi
}%
\providecommand \@ifx [1]{%
 \ifx #1\expandafter \@firstoftwo
 \else \expandafter \@secondoftwo
 \fi
}%
\providecommand \natexlab [1]{#1}%
\providecommand \enquote  [1]{``#1''}%
\providecommand \bibnamefont  [1]{#1}%
\providecommand \bibfnamefont [1]{#1}%
\providecommand \citenamefont [1]{#1}%
\providecommand \href@noop [0]{\@secondoftwo}%
\providecommand \href [0]{\begingroup \@sanitize@url \@href}%
\providecommand \@href[1]{\@@startlink{#1}\@@href}%
\providecommand \@@href[1]{\endgroup#1\@@endlink}%
\providecommand \@sanitize@url [0]{\catcode `\\12\catcode `\$12\catcode `\&12\catcode `\#12\catcode `\^12\catcode `\_12\catcode `\%12\relax}%
\providecommand \@@startlink[1]{}%
\providecommand \@@endlink[0]{}%
\providecommand \url  [0]{\begingroup\@sanitize@url \@url }%
\providecommand \@url [1]{\endgroup\@href {#1}{\urlprefix }}%
\providecommand \urlprefix  [0]{URL }%
\providecommand \Eprint [0]{\href }%
\providecommand \doibase [0]{https://doi.org/}%
\providecommand \selectlanguage [0]{\@gobble}%
\providecommand \bibinfo  [0]{\@secondoftwo}%
\providecommand \bibfield  [0]{\@secondoftwo}%
\providecommand \translation [1]{[#1]}%
\providecommand \BibitemOpen [0]{}%
\providecommand \bibitemStop [0]{}%
\providecommand \bibitemNoStop [0]{.\EOS\space}%
\providecommand \EOS [0]{\spacefactor3000\relax}%
\providecommand \BibitemShut  [1]{\csname bibitem#1\endcsname}%
\let\auto@bib@innerbib\@empty
\bibitem [{\citenamefont {Cassak}\ \emph {et~al.}(2017)\citenamefont {Cassak}, \citenamefont {Liu},\ and\ \citenamefont {Shay}}]{Cassak2017}%
  \BibitemOpen
  \bibfield  {author} {\bibinfo {author} {\bibfnamefont {P.~A.}\ \bibnamefont {Cassak}}, \bibinfo {author} {\bibfnamefont {Y.-H.}\ \bibnamefont {Liu}},\ and\ \bibinfo {author} {\bibfnamefont {M.~A.}\ \bibnamefont {Shay}},\ }\bibfield  {title} {\bibinfo {title} {A review of the 0.1 reconnection rate problem},\ }\href {https://doi.org/10.1017/S002237781700066X} {\bibfield  {journal} {\bibinfo  {journal} {Journal of Plasma Physics}\ }\textbf {\bibinfo {volume} {83}},\ \bibinfo {pages} {715830501} (\bibinfo {year} {2017})}\BibitemShut {NoStop}%
\bibitem [{\citenamefont {Parker}(1957)}]{Parker1957}%
  \BibitemOpen
  \bibfield  {author} {\bibinfo {author} {\bibfnamefont {E.~N.}\ \bibnamefont {Parker}},\ }\bibfield  {title} {\bibinfo {title} {Sweet's mechanism for merging magnetic fields in conducting fluids},\ }\href {https://doi.org/10.1029/JZ062i004p00509} {\bibfield  {journal} {\bibinfo  {journal} {Journal of Geophysical Research}\ }\textbf {\bibinfo {volume} {62}},\ \bibinfo {pages} {509} (\bibinfo {year} {1957})}\BibitemShut {NoStop}%
\bibitem [{\citenamefont {Sweet}(1958)}]{Sweet1958}%
  \BibitemOpen
  \bibfield  {author} {\bibinfo {author} {\bibfnamefont {P.~A.}\ \bibnamefont {Sweet}},\ }\bibfield  {title} {\bibinfo {title} {The neutral point theory of solar flares},\ }in\ \href@noop {} {\emph {\bibinfo {booktitle} {Electromagnetic Phenomena in Cosmical Physics}}},\ \bibinfo {series} {IAU Symposium}, Vol.~\bibinfo {volume} {6},\ \bibinfo {editor} {edited by\ \bibinfo {editor} {\bibfnamefont {B.}~\bibnamefont {Lehnert}}}\ (\bibinfo  {publisher} {Cambridge University Press},\ \bibinfo {year} {1958})\ p.\ \bibinfo {pages} {123}\BibitemShut {NoStop}%
\bibitem [{\citenamefont {Petschek}(1964)}]{Petschek1964}%
  \BibitemOpen
  \bibfield  {author} {\bibinfo {author} {\bibfnamefont {H.~E.}\ \bibnamefont {Petschek}},\ }\bibfield  {title} {\bibinfo {title} {Magnetic field annihilation},\ }\href@noop {} {\bibfield  {journal} {\bibinfo  {journal} {NASA Special Publication}\ }\textbf {\bibinfo {volume} {50}},\ \bibinfo {pages} {425} (\bibinfo {year} {1964})}\BibitemShut {NoStop}%
\bibitem [{\citenamefont {Biskamp}(1986)}]{Biskamp1986}%
  \BibitemOpen
  \bibfield  {author} {\bibinfo {author} {\bibfnamefont {D.}~\bibnamefont {Biskamp}},\ }\bibfield  {title} {\bibinfo {title} {Magnetic reconnection via current sheets},\ }\href {https://doi.org/10.1063/1.865670} {\bibfield  {journal} {\bibinfo  {journal} {Physics of Fluids}\ }\textbf {\bibinfo {volume} {29}},\ \bibinfo {pages} {1520} (\bibinfo {year} {1986})}\BibitemShut {NoStop}%
\bibitem [{\citenamefont {Ugai}\ and\ \citenamefont {Tsuda}(1977)}]{Ugai1977}%
  \BibitemOpen
  \bibfield  {author} {\bibinfo {author} {\bibfnamefont {M.}~\bibnamefont {Ugai}}\ and\ \bibinfo {author} {\bibfnamefont {T.}~\bibnamefont {Tsuda}},\ }\bibfield  {title} {\bibinfo {title} {Magnetic field-line reconnexion by localized enhancement of resistivity. part 1. evolution in a compressible {MHD} fluid},\ }\href {https://doi.org/10.1017/S0022377800020663} {\bibfield  {journal} {\bibinfo  {journal} {Journal of Plasma Physics}\ }\textbf {\bibinfo {volume} {17}},\ \bibinfo {pages} {337} (\bibinfo {year} {1977})}\BibitemShut {NoStop}%
\bibitem [{\citenamefont {Kulsrud}(2001)}]{Kulsrud2001}%
  \BibitemOpen
  \bibfield  {author} {\bibinfo {author} {\bibfnamefont {R.~M.}\ \bibnamefont {Kulsrud}},\ }\bibfield  {title} {\bibinfo {title} {Magnetic reconnection: {Sweet--Parker} versus {Petschek}},\ }\href {https://doi.org/10.1186/BF03353251} {\bibfield  {journal} {\bibinfo  {journal} {Earth, Planets and Space}\ }\textbf {\bibinfo {volume} {53}},\ \bibinfo {pages} {417} (\bibinfo {year} {2001})}\BibitemShut {NoStop}%
\bibitem [{\citenamefont {Shay}\ \emph {et~al.}(1998)\citenamefont {Shay}, \citenamefont {Drake}, \citenamefont {Denton},\ and\ \citenamefont {Biskamp}}]{Shay1998}%
  \BibitemOpen
  \bibfield  {author} {\bibinfo {author} {\bibfnamefont {M.~A.}\ \bibnamefont {Shay}}, \bibinfo {author} {\bibfnamefont {J.~F.}\ \bibnamefont {Drake}}, \bibinfo {author} {\bibfnamefont {R.~E.}\ \bibnamefont {Denton}},\ and\ \bibinfo {author} {\bibfnamefont {D.}~\bibnamefont {Biskamp}},\ }\bibfield  {title} {\bibinfo {title} {Structure of the dissipation region during collisionless magnetic reconnection},\ }\href {https://doi.org/10.1029/97JA03528} {\bibfield  {journal} {\bibinfo  {journal} {Journal of Geophysical Research: Space Physics}\ }\textbf {\bibinfo {volume} {103}},\ \bibinfo {pages} {9165} (\bibinfo {year} {1998})}\BibitemShut {NoStop}%
\bibitem [{\citenamefont {Birn}\ \emph {et~al.}(2001)\citenamefont {Birn}, \citenamefont {Drake}, \citenamefont {Shay}, \citenamefont {Rogers}, \citenamefont {Denton}, \citenamefont {Hesse}, \citenamefont {Kuznetsova}, \citenamefont {Ma}, \citenamefont {Bhattacharjee}, \citenamefont {Otto},\ and\ \citenamefont {Pritchett}}]{Birn2001}%
  \BibitemOpen
  \bibfield  {author} {\bibinfo {author} {\bibfnamefont {J.}~\bibnamefont {Birn}}, \bibinfo {author} {\bibfnamefont {J.~F.}\ \bibnamefont {Drake}}, \bibinfo {author} {\bibfnamefont {M.~A.}\ \bibnamefont {Shay}}, \bibinfo {author} {\bibfnamefont {B.~N.}\ \bibnamefont {Rogers}}, \bibinfo {author} {\bibfnamefont {R.~E.}\ \bibnamefont {Denton}}, \bibinfo {author} {\bibfnamefont {M.}~\bibnamefont {Hesse}}, \bibinfo {author} {\bibfnamefont {M.}~\bibnamefont {Kuznetsova}}, \bibinfo {author} {\bibfnamefont {Z.~W.}\ \bibnamefont {Ma}}, \bibinfo {author} {\bibfnamefont {A.}~\bibnamefont {Bhattacharjee}}, \bibinfo {author} {\bibfnamefont {A.}~\bibnamefont {Otto}},\ and\ \bibinfo {author} {\bibfnamefont {P.~L.}\ \bibnamefont {Pritchett}},\ }\bibfield  {title} {\bibinfo {title} {Geospace environmental modeling ({GEM}) magnetic reconnection challenge},\ }\href {https://doi.org/10.1029/1999JA900449} {\bibfield  {journal} {\bibinfo  {journal} {Journal of Geophysical Research: Space Physics}\ }\textbf {\bibinfo {volume} {106}},\ \bibinfo {pages} {3715} (\bibinfo {year} {2001})}\BibitemShut {NoStop}%
\bibitem [{\citenamefont {Bessho}\ and\ \citenamefont {Bhattacharjee}(2005)}]{Bessho2005}%
  \BibitemOpen
  \bibfield  {author} {\bibinfo {author} {\bibfnamefont {N.}~\bibnamefont {Bessho}}\ and\ \bibinfo {author} {\bibfnamefont {A.}~\bibnamefont {Bhattacharjee}},\ }\bibfield  {title} {\bibinfo {title} {Collisionless reconnection in an electron-positron plasma},\ }\href {https://doi.org/10.1103/PhysRevLett.95.245001} {\bibfield  {journal} {\bibinfo  {journal} {Physical Review Letters}\ }\textbf {\bibinfo {volume} {95}},\ \bibinfo {pages} {245001} (\bibinfo {year} {2005})}\BibitemShut {NoStop}%
\bibitem [{\citenamefont {Bessho}\ and\ \citenamefont {Bhattacharjee}(2007)}]{Bessho2007}%
  \BibitemOpen
  \bibfield  {author} {\bibinfo {author} {\bibfnamefont {N.}~\bibnamefont {Bessho}}\ and\ \bibinfo {author} {\bibfnamefont {A.}~\bibnamefont {Bhattacharjee}},\ }\bibfield  {title} {\bibinfo {title} {Fast collisionless reconnection in electron-positron plasmas},\ }\href {https://doi.org/10.1063/1.2714020} {\bibfield  {journal} {\bibinfo  {journal} {Physics of Plasmas}\ }\textbf {\bibinfo {volume} {14}},\ \bibinfo {pages} {056503} (\bibinfo {year} {2007})}\BibitemShut {NoStop}%
\bibitem [{\citenamefont {Hesse}\ \emph {et~al.}(2009)\citenamefont {Hesse}, \citenamefont {Zenitani}, \citenamefont {Kuznetsova},\ and\ \citenamefont {Klimas}}]{Hesse2009}%
  \BibitemOpen
  \bibfield  {author} {\bibinfo {author} {\bibfnamefont {M.}~\bibnamefont {Hesse}}, \bibinfo {author} {\bibfnamefont {S.}~\bibnamefont {Zenitani}}, \bibinfo {author} {\bibfnamefont {M.}~\bibnamefont {Kuznetsova}},\ and\ \bibinfo {author} {\bibfnamefont {A.}~\bibnamefont {Klimas}},\ }\bibfield  {title} {\bibinfo {title} {A simple, analytical model of collisionless magnetic reconnection in a pair plasma},\ }\href {https://doi.org/10.1063/1.3246005} {\bibfield  {journal} {\bibinfo  {journal} {Physics of Plasmas}\ }\textbf {\bibinfo {volume} {16}},\ \bibinfo {pages} {102106} (\bibinfo {year} {2009})}\BibitemShut {NoStop}%
\bibitem [{\citenamefont {Daughton}\ \emph {et~al.}(2006)\citenamefont {Daughton}, \citenamefont {Scudder},\ and\ \citenamefont {Karimabadi}}]{Daughton2006}%
  \BibitemOpen
  \bibfield  {author} {\bibinfo {author} {\bibfnamefont {W.}~\bibnamefont {Daughton}}, \bibinfo {author} {\bibfnamefont {J.}~\bibnamefont {Scudder}},\ and\ \bibinfo {author} {\bibfnamefont {H.}~\bibnamefont {Karimabadi}},\ }\bibfield  {title} {\bibinfo {title} {Fully kinetic simulations of undriven magnetic reconnection with open boundary conditions},\ }\href {https://doi.org/10.1063/1.2218817} {\bibfield  {journal} {\bibinfo  {journal} {Physics of Plasmas}\ }\textbf {\bibinfo {volume} {13}},\ \bibinfo {pages} {072101} (\bibinfo {year} {2006})}\BibitemShut {NoStop}%
\bibitem [{\citenamefont {Daughton}\ and\ \citenamefont {Karimabadi}(2007)}]{Daughton2007}%
  \BibitemOpen
  \bibfield  {author} {\bibinfo {author} {\bibfnamefont {W.}~\bibnamefont {Daughton}}\ and\ \bibinfo {author} {\bibfnamefont {H.}~\bibnamefont {Karimabadi}},\ }\bibfield  {title} {\bibinfo {title} {Collisionless magnetic reconnection in large-scale electron-positron plasmas},\ }\href {https://doi.org/10.1063/1.2749494} {\bibfield  {journal} {\bibinfo  {journal} {Physics of Plasmas}\ }\textbf {\bibinfo {volume} {14}},\ \bibinfo {pages} {072303} (\bibinfo {year} {2007})}\BibitemShut {NoStop}%
\bibitem [{\citenamefont {Swisdak}\ \emph {et~al.}(2008)\citenamefont {Swisdak}, \citenamefont {Liu},\ and\ \citenamefont {Drake}}]{Swisdak2008}%
  \BibitemOpen
  \bibfield  {author} {\bibinfo {author} {\bibfnamefont {M.}~\bibnamefont {Swisdak}}, \bibinfo {author} {\bibfnamefont {Y.-H.}\ \bibnamefont {Liu}},\ and\ \bibinfo {author} {\bibfnamefont {J.~F.}\ \bibnamefont {Drake}},\ }\bibfield  {title} {\bibinfo {title} {Development of a turbulent outflow during electron-positron magnetic reconnection},\ }\href {https://doi.org/10.1086/588088} {\bibfield  {journal} {\bibinfo  {journal} {The Astrophysical Journal}\ }\textbf {\bibinfo {volume} {680}},\ \bibinfo {pages} {999} (\bibinfo {year} {2008})}\BibitemShut {NoStop}%
\bibitem [{\citenamefont {Loureiro}\ \emph {et~al.}(2007)\citenamefont {Loureiro}, \citenamefont {Schekochihin},\ and\ \citenamefont {Cowley}}]{Loureiro2007}%
  \BibitemOpen
  \bibfield  {author} {\bibinfo {author} {\bibfnamefont {N.~F.}\ \bibnamefont {Loureiro}}, \bibinfo {author} {\bibfnamefont {A.~A.}\ \bibnamefont {Schekochihin}},\ and\ \bibinfo {author} {\bibfnamefont {S.~C.}\ \bibnamefont {Cowley}},\ }\bibfield  {title} {\bibinfo {title} {Instability of current sheets and formation of plasmoid chains},\ }\href {https://doi.org/10.1063/1.2783986} {\bibfield  {journal} {\bibinfo  {journal} {Physics of Plasmas}\ }\textbf {\bibinfo {volume} {14}},\ \bibinfo {pages} {100703} (\bibinfo {year} {2007})}\BibitemShut {NoStop}%
\bibitem [{\citenamefont {Bhattacharjee}\ \emph {et~al.}(2009)\citenamefont {Bhattacharjee}, \citenamefont {Huang}, \citenamefont {Yang},\ and\ \citenamefont {Rogers}}]{Bhattacharjee2009}%
  \BibitemOpen
  \bibfield  {author} {\bibinfo {author} {\bibfnamefont {A.}~\bibnamefont {Bhattacharjee}}, \bibinfo {author} {\bibfnamefont {Y.-M.}\ \bibnamefont {Huang}}, \bibinfo {author} {\bibfnamefont {H.}~\bibnamefont {Yang}},\ and\ \bibinfo {author} {\bibfnamefont {B.}~\bibnamefont {Rogers}},\ }\bibfield  {title} {\bibinfo {title} {Fast reconnection in high-{Lundquist}-number plasmas due to the plasmoid instability},\ }\href {https://doi.org/10.1063/1.3264103} {\bibfield  {journal} {\bibinfo  {journal} {Physics of Plasmas}\ }\textbf {\bibinfo {volume} {16}},\ \bibinfo {pages} {112102} (\bibinfo {year} {2009})}\BibitemShut {NoStop}%
\bibitem [{\citenamefont {Uzdensky}\ \emph {et~al.}(2010)\citenamefont {Uzdensky}, \citenamefont {Loureiro},\ and\ \citenamefont {Schekochihin}}]{Uzdensky2010}%
  \BibitemOpen
  \bibfield  {author} {\bibinfo {author} {\bibfnamefont {D.~A.}\ \bibnamefont {Uzdensky}}, \bibinfo {author} {\bibfnamefont {N.~F.}\ \bibnamefont {Loureiro}},\ and\ \bibinfo {author} {\bibfnamefont {A.~A.}\ \bibnamefont {Schekochihin}},\ }\bibfield  {title} {\bibinfo {title} {Fast magnetic reconnection in the plasmoid-dominated regime},\ }\href {https://doi.org/10.1103/PhysRevLett.105.235002} {\bibfield  {journal} {\bibinfo  {journal} {Physical Review Letters}\ }\textbf {\bibinfo {volume} {105}},\ \bibinfo {pages} {235002} (\bibinfo {year} {2010})}\BibitemShut {NoStop}%
\bibitem [{\citenamefont {Sironi}\ and\ \citenamefont {Spitkovsky}(2014)}]{Sironi2014}%
  \BibitemOpen
  \bibfield  {author} {\bibinfo {author} {\bibfnamefont {L.}~\bibnamefont {Sironi}}\ and\ \bibinfo {author} {\bibfnamefont {A.}~\bibnamefont {Spitkovsky}},\ }\bibfield  {title} {\bibinfo {title} {Relativistic reconnection: an efficient source of non-thermal particles},\ }\href {https://doi.org/10.1088/2041-8205/783/1/L21} {\bibfield  {journal} {\bibinfo  {journal} {The Astrophysical Journal Letters}\ }\textbf {\bibinfo {volume} {783}},\ \bibinfo {pages} {L21} (\bibinfo {year} {2014})}\BibitemShut {NoStop}%
\bibitem [{\citenamefont {Sironi}\ \emph {et~al.}(2016)\citenamefont {Sironi}, \citenamefont {Giannios},\ and\ \citenamefont {Petropoulou}}]{Sironi2016}%
  \BibitemOpen
  \bibfield  {author} {\bibinfo {author} {\bibfnamefont {L.}~\bibnamefont {Sironi}}, \bibinfo {author} {\bibfnamefont {D.}~\bibnamefont {Giannios}},\ and\ \bibinfo {author} {\bibfnamefont {M.}~\bibnamefont {Petropoulou}},\ }\bibfield  {title} {\bibinfo {title} {Plasmoids in relativistic reconnection, from birth to adulthood: first they grow, then they go},\ }\href {https://doi.org/10.1093/mnras/stw1620} {\bibfield  {journal} {\bibinfo  {journal} {Monthly Notices of the Royal Astronomical Society}\ }\textbf {\bibinfo {volume} {462}},\ \bibinfo {pages} {48} (\bibinfo {year} {2016})}\BibitemShut {NoStop}%
\bibitem [{\citenamefont {Ortu{\~n}o-Mac{\'i}as}\ and\ \citenamefont {Nalewajko}(2020)}]{OrtunoMacias2020}%
  \BibitemOpen
  \bibfield  {author} {\bibinfo {author} {\bibfnamefont {J.}~\bibnamefont {Ortu{\~n}o-Mac{\'i}as}}\ and\ \bibinfo {author} {\bibfnamefont {K.}~\bibnamefont {Nalewajko}},\ }\bibfield  {title} {\bibinfo {title} {Radiative kinetic simulations of steady-state relativistic plasmoid magnetic reconnection},\ }\href {https://doi.org/10.1093/mnras/staa1899} {\bibfield  {journal} {\bibinfo  {journal} {Monthly Notices of the Royal Astronomical Society}\ }\textbf {\bibinfo {volume} {497}},\ \bibinfo {pages} {1365} (\bibinfo {year} {2020})}\BibitemShut {NoStop}%
\bibitem [{\citenamefont {Robbins}\ and\ \citenamefont {Spitkovsky}(2025)}]{Robbins2025}%
  \BibitemOpen
  \bibfield  {author} {\bibinfo {author} {\bibfnamefont {A.}~\bibnamefont {Robbins}}\ and\ \bibinfo {author} {\bibfnamefont {A.}~\bibnamefont {Spitkovsky}},\ }\bibfield  {title} {\bibinfo {title} {Transition to {Petschek} reconnection in subrelativistic pair plasmas: Implications for particle acceleration},\ }\href {https://doi.org/10.3847/1538-4357/ae1618} {\bibfield  {journal} {\bibinfo  {journal} {The Astrophysical Journal}\ }\textbf {\bibinfo {volume} {995}},\ \bibinfo {pages} {163} (\bibinfo {year} {2025})}\BibitemShut {NoStop}%
\bibitem [{\citenamefont {Speiser}(1965)}]{Speiser1965}%
  \BibitemOpen
  \bibfield  {author} {\bibinfo {author} {\bibfnamefont {T.~W.}\ \bibnamefont {Speiser}},\ }\bibfield  {title} {\bibinfo {title} {Particle trajectories in model current sheets, 1, analytical solutions},\ }\href {https://doi.org/10.1029/JZ070i017p04219} {\bibfield  {journal} {\bibinfo  {journal} {Journal of Geophysical Research}\ }\textbf {\bibinfo {volume} {70}},\ \bibinfo {pages} {4219} (\bibinfo {year} {1965})}\BibitemShut {NoStop}%
\bibitem [{\citenamefont {Speiser}(1970)}]{Speiser1970}%
  \BibitemOpen
  \bibfield  {author} {\bibinfo {author} {\bibfnamefont {T.~W.}\ \bibnamefont {Speiser}},\ }\bibfield  {title} {\bibinfo {title} {Conductivity without collisions or noise},\ }\href {https://doi.org/10.1016/0032-0633(70)90136-4} {\bibfield  {journal} {\bibinfo  {journal} {Planetary and Space Science}\ }\textbf {\bibinfo {volume} {18}},\ \bibinfo {pages} {613} (\bibinfo {year} {1970})}\BibitemShut {NoStop}%
\bibitem [{\citenamefont {Goodbred}\ and\ \citenamefont {Spitkovsky}(2026)}]{Goodbred2026}%
  \BibitemOpen
  \bibfield  {author} {\bibinfo {author} {\bibfnamefont {M.}~\bibnamefont {Goodbred}}\ and\ \bibinfo {author} {\bibfnamefont {A.}~\bibnamefont {Spitkovsky}},\ }\href@noop {} {\bibinfo {title} {The birth of plasmoids: {Weibel}-driven secondary tearing and its suppression in relativistic reconnection}} (\bibinfo {year} {2026}),\ \Eprint {https://arxiv.org/abs/2611.02126} {arXiv:2611.02126 [astro-ph.HE]} \BibitemShut {NoStop}%
\bibitem [{\citenamefont {Lyons}\ and\ \citenamefont {Speiser}(1985)}]{Lyons1985}%
  \BibitemOpen
  \bibfield  {author} {\bibinfo {author} {\bibfnamefont {L.~R.}\ \bibnamefont {Lyons}}\ and\ \bibinfo {author} {\bibfnamefont {T.~W.}\ \bibnamefont {Speiser}},\ }\bibfield  {title} {\bibinfo {title} {Ohm's law for a current sheet},\ }\href {https://doi.org/10.1029/JA090iA09p08543} {\bibfield  {journal} {\bibinfo  {journal} {Journal of Geophysical Research}\ }\textbf {\bibinfo {volume} {90}},\ \bibinfo {pages} {8543} (\bibinfo {year} {1985})}\BibitemShut {NoStop}%
\bibitem [{\citenamefont {Selvi}\ \emph {et~al.}(2023)\citenamefont {Selvi}, \citenamefont {Porth}, \citenamefont {Ripperda}, \citenamefont {Bacchini}, \citenamefont {Sironi},\ and\ \citenamefont {Keppens}}]{Selvi2023}%
  \BibitemOpen
  \bibfield  {author} {\bibinfo {author} {\bibfnamefont {S.}~\bibnamefont {Selvi}}, \bibinfo {author} {\bibfnamefont {O.}~\bibnamefont {Porth}}, \bibinfo {author} {\bibfnamefont {B.}~\bibnamefont {Ripperda}}, \bibinfo {author} {\bibfnamefont {F.}~\bibnamefont {Bacchini}}, \bibinfo {author} {\bibfnamefont {L.}~\bibnamefont {Sironi}},\ and\ \bibinfo {author} {\bibfnamefont {R.}~\bibnamefont {Keppens}},\ }\bibfield  {title} {\bibinfo {title} {Effective resistivity in relativistic collisionless reconnection},\ }\href {https://doi.org/10.3847/1538-4357/acd0b0} {\bibfield  {journal} {\bibinfo  {journal} {The Astrophysical Journal}\ }\textbf {\bibinfo {volume} {950}},\ \bibinfo {pages} {169} (\bibinfo {year} {2023})}\BibitemShut {NoStop}%
\bibitem [{\citenamefont {Vasyliunas}(1975)}]{Vasyliunas1975}%
  \BibitemOpen
  \bibfield  {author} {\bibinfo {author} {\bibfnamefont {V.~M.}\ \bibnamefont {Vasyliunas}},\ }\bibfield  {title} {\bibinfo {title} {Theoretical models of magnetic field line merging, 1},\ }\href {https://doi.org/10.1029/RG013i001p00303} {\bibfield  {journal} {\bibinfo  {journal} {Reviews of Geophysics}\ }\textbf {\bibinfo {volume} {13}},\ \bibinfo {pages} {303} (\bibinfo {year} {1975})}\BibitemShut {NoStop}%
\bibitem [{\citenamefont {Kuznetsova}\ \emph {et~al.}(1998)\citenamefont {Kuznetsova}, \citenamefont {Hesse},\ and\ \citenamefont {Winske}}]{Kuznetsova1998}%
  \BibitemOpen
  \bibfield  {author} {\bibinfo {author} {\bibfnamefont {M.~M.}\ \bibnamefont {Kuznetsova}}, \bibinfo {author} {\bibfnamefont {M.}~\bibnamefont {Hesse}},\ and\ \bibinfo {author} {\bibfnamefont {D.}~\bibnamefont {Winske}},\ }\bibfield  {title} {\bibinfo {title} {Kinetic quasi-viscous and bulk flow inertia effects in collisionless magnetotail reconnection},\ }\href {https://doi.org/10.1029/97JA02699} {\bibfield  {journal} {\bibinfo  {journal} {Journal of Geophysical Research: Space Physics}\ }\textbf {\bibinfo {volume} {103}},\ \bibinfo {pages} {199} (\bibinfo {year} {1998})}\BibitemShut {NoStop}%
\bibitem [{\citenamefont {Hesse}\ \emph {et~al.}(1999)\citenamefont {Hesse}, \citenamefont {Schindler}, \citenamefont {Birn},\ and\ \citenamefont {Kuznetsova}}]{Hesse1999}%
  \BibitemOpen
  \bibfield  {author} {\bibinfo {author} {\bibfnamefont {M.}~\bibnamefont {Hesse}}, \bibinfo {author} {\bibfnamefont {K.}~\bibnamefont {Schindler}}, \bibinfo {author} {\bibfnamefont {J.}~\bibnamefont {Birn}},\ and\ \bibinfo {author} {\bibfnamefont {M.}~\bibnamefont {Kuznetsova}},\ }\bibfield  {title} {\bibinfo {title} {The diffusion region in collisionless magnetic reconnection},\ }\href {https://doi.org/10.1063/1.873436} {\bibfield  {journal} {\bibinfo  {journal} {Physics of Plasmas}\ }\textbf {\bibinfo {volume} {6}},\ \bibinfo {pages} {1781} (\bibinfo {year} {1999})}\BibitemShut {NoStop}%
\bibitem [{\citenamefont {Liu}\ \emph {et~al.}(2017)\citenamefont {Liu}, \citenamefont {Hesse}, \citenamefont {Guo}, \citenamefont {Daughton}, \citenamefont {Li}, \citenamefont {Cassak},\ and\ \citenamefont {Shay}}]{Liu2017}%
  \BibitemOpen
  \bibfield  {author} {\bibinfo {author} {\bibfnamefont {Y.-H.}\ \bibnamefont {Liu}}, \bibinfo {author} {\bibfnamefont {M.}~\bibnamefont {Hesse}}, \bibinfo {author} {\bibfnamefont {F.}~\bibnamefont {Guo}}, \bibinfo {author} {\bibfnamefont {W.}~\bibnamefont {Daughton}}, \bibinfo {author} {\bibfnamefont {H.}~\bibnamefont {Li}}, \bibinfo {author} {\bibfnamefont {P.~A.}\ \bibnamefont {Cassak}},\ and\ \bibinfo {author} {\bibfnamefont {M.~A.}\ \bibnamefont {Shay}},\ }\bibfield  {title} {\bibinfo {title} {Why does steady-state magnetic reconnection have a maximum local rate of $\sim 0.1$?},\ }\href {https://doi.org/10.1103/PhysRevLett.118.085101} {\bibfield  {journal} {\bibinfo  {journal} {Physical Review Letters}\ }\textbf {\bibinfo {volume} {118}},\ \bibinfo {pages} {085101} (\bibinfo {year} {2017})}\BibitemShut {NoStop}%
\bibitem [{\citenamefont {Liu}\ \emph {et~al.}(2020)\citenamefont {Liu}, \citenamefont {Lin}, \citenamefont {Hesse}, \citenamefont {Guo}, \citenamefont {Li}, \citenamefont {Zhang},\ and\ \citenamefont {Peery}}]{Liu2020}%
  \BibitemOpen
  \bibfield  {author} {\bibinfo {author} {\bibfnamefont {Y.-H.}\ \bibnamefont {Liu}}, \bibinfo {author} {\bibfnamefont {S.-C.}\ \bibnamefont {Lin}}, \bibinfo {author} {\bibfnamefont {M.}~\bibnamefont {Hesse}}, \bibinfo {author} {\bibfnamefont {F.}~\bibnamefont {Guo}}, \bibinfo {author} {\bibfnamefont {X.}~\bibnamefont {Li}}, \bibinfo {author} {\bibfnamefont {H.}~\bibnamefont {Zhang}},\ and\ \bibinfo {author} {\bibfnamefont {S.}~\bibnamefont {Peery}},\ }\bibfield  {title} {\bibinfo {title} {The critical role of collisionless plasma energization on the structure of relativistic magnetic reconnection},\ }\href {https://doi.org/10.3847/2041-8213/ab7d3f} {\bibfield  {journal} {\bibinfo  {journal} {The Astrophysical Journal Letters}\ }\textbf {\bibinfo {volume} {892}},\ \bibinfo {pages} {L13} (\bibinfo {year} {2020})}\BibitemShut {NoStop}%
\bibitem [{\citenamefont {Liu}\ \emph {et~al.}(2022)\citenamefont {Liu}, \citenamefont {Cassak}, \citenamefont {Li}, \citenamefont {Hesse}, \citenamefont {Lin},\ and\ \citenamefont {Genestreti}}]{Liu2022}%
  \BibitemOpen
  \bibfield  {author} {\bibinfo {author} {\bibfnamefont {Y.-H.}\ \bibnamefont {Liu}}, \bibinfo {author} {\bibfnamefont {P.~A.}\ \bibnamefont {Cassak}}, \bibinfo {author} {\bibfnamefont {X.}~\bibnamefont {Li}}, \bibinfo {author} {\bibfnamefont {M.}~\bibnamefont {Hesse}}, \bibinfo {author} {\bibfnamefont {S.-C.}\ \bibnamefont {Lin}},\ and\ \bibinfo {author} {\bibfnamefont {K.}~\bibnamefont {Genestreti}},\ }\bibfield  {title} {\bibinfo {title} {First-principles theory of the rate of magnetic reconnection in magnetospheric and solar plasmas},\ }\href {https://doi.org/10.1038/s42005-022-00854-x} {\bibfield  {journal} {\bibinfo  {journal} {Communications Physics}\ }\textbf {\bibinfo {volume} {5}},\ \bibinfo {pages} {97} (\bibinfo {year} {2022})}\BibitemShut {NoStop}%
\bibitem [{\citenamefont {Goodbred}\ and\ \citenamefont {Liu}(2022)}]{Goodbred2022}%
  \BibitemOpen
  \bibfield  {author} {\bibinfo {author} {\bibfnamefont {M.}~\bibnamefont {Goodbred}}\ and\ \bibinfo {author} {\bibfnamefont {Y.-H.}\ \bibnamefont {Liu}},\ }\bibfield  {title} {\bibinfo {title} {First-principles theory of the relativistic magnetic reconnection rate in astrophysical pair plasmas},\ }\href {https://doi.org/10.1103/PhysRevLett.129.265101} {\bibfield  {journal} {\bibinfo  {journal} {Physical Review Letters}\ }\textbf {\bibinfo {volume} {129}},\ \bibinfo {pages} {265101} (\bibinfo {year} {2022})}\BibitemShut {NoStop}%
\bibitem [{\citenamefont {Bulanov}\ and\ \citenamefont {Sasorov}(1975)}]{Bulanov1975}%
  \BibitemOpen
  \bibfield  {author} {\bibinfo {author} {\bibfnamefont {S.~V.}\ \bibnamefont {Bulanov}}\ and\ \bibinfo {author} {\bibfnamefont {P.~V.}\ \bibnamefont {Sasorov}},\ }\bibfield  {title} {\bibinfo {title} {Energy spectrum of particles accelerated in the neighborhood of a line of zero magnetic field},\ }\href {https://ui.adsabs.harvard.edu/abs/1976SvA....19..464B} {\bibfield  {journal} {\bibinfo  {journal} {Soviet Astronomy}\ }\textbf {\bibinfo {volume} {19}},\ \bibinfo {pages} {464} (\bibinfo {year} {1975})}\BibitemShut {NoStop}%
\bibitem [{\citenamefont {Zenitani}\ and\ \citenamefont {Hoshino}(2001)}]{Zenitani2001}%
  \BibitemOpen
  \bibfield  {author} {\bibinfo {author} {\bibfnamefont {S.}~\bibnamefont {Zenitani}}\ and\ \bibinfo {author} {\bibfnamefont {M.}~\bibnamefont {Hoshino}},\ }\bibfield  {title} {\bibinfo {title} {The generation of nonthermal particles in relativistic magnetic reconnection of pair plasmas},\ }\href {https://doi.org/10.1086/337972} {\bibfield  {journal} {\bibinfo  {journal} {The Astrophysical Journal Letters}\ }\textbf {\bibinfo {volume} {562}},\ \bibinfo {pages} {L63} (\bibinfo {year} {2001})}\BibitemShut {NoStop}%
\bibitem [{\citenamefont {Larrabee}\ \emph {et~al.}(2003)\citenamefont {Larrabee}, \citenamefont {Lovelace},\ and\ \citenamefont {Romanova}}]{Larrabee2003}%
  \BibitemOpen
  \bibfield  {author} {\bibinfo {author} {\bibfnamefont {D.~A.}\ \bibnamefont {Larrabee}}, \bibinfo {author} {\bibfnamefont {R.~V.~E.}\ \bibnamefont {Lovelace}},\ and\ \bibinfo {author} {\bibfnamefont {M.~M.}\ \bibnamefont {Romanova}},\ }\bibfield  {title} {\bibinfo {title} {Lepton acceleration by relativistic collisionless magnetic reconnection},\ }\href {https://doi.org/10.1086/367640} {\bibfield  {journal} {\bibinfo  {journal} {The Astrophysical Journal}\ }\textbf {\bibinfo {volume} {586}},\ \bibinfo {pages} {72} (\bibinfo {year} {2003})}\BibitemShut {NoStop}%
\bibitem [{\citenamefont {Bessho}\ and\ \citenamefont {Bhattacharjee}(2012)}]{Bessho2012}%
  \BibitemOpen
  \bibfield  {author} {\bibinfo {author} {\bibfnamefont {N.}~\bibnamefont {Bessho}}\ and\ \bibinfo {author} {\bibfnamefont {A.}~\bibnamefont {Bhattacharjee}},\ }\bibfield  {title} {\bibinfo {title} {Fast magnetic reconnection and particle acceleration in relativistic low-density electron--positron plasmas without guide field},\ }\href {https://doi.org/10.1088/0004-637X/750/2/129} {\bibfield  {journal} {\bibinfo  {journal} {The Astrophysical Journal}\ }\textbf {\bibinfo {volume} {750}},\ \bibinfo {pages} {129} (\bibinfo {year} {2012})}\BibitemShut {NoStop}%
\bibitem [{\citenamefont {Weibel}(1959)}]{Weibel1959}%
  \BibitemOpen
  \bibfield  {author} {\bibinfo {author} {\bibfnamefont {E.~S.}\ \bibnamefont {Weibel}},\ }\bibfield  {title} {\bibinfo {title} {Spontaneously growing transverse waves in a plasma due to an anisotropic velocity distribution},\ }\href {https://doi.org/10.1103/PhysRevLett.2.83} {\bibfield  {journal} {\bibinfo  {journal} {Physical Review Letters}\ }\textbf {\bibinfo {volume} {2}},\ \bibinfo {pages} {83} (\bibinfo {year} {1959})}\BibitemShut {NoStop}%
\bibitem [{\citenamefont {Coppi}(1983)}]{Coppi1983}%
  \BibitemOpen
  \bibfield  {author} {\bibinfo {author} {\bibfnamefont {B.}~\bibnamefont {Coppi}},\ }\bibfield  {title} {\bibinfo {title} {Magnetic reconnection driven by velocity space instabilities},\ }\href {https://ui.adsabs.harvard.edu/abs/1983ApJ...273L.101C} {\bibfield  {journal} {\bibinfo  {journal} {The Astrophysical Journal Letters}\ }\textbf {\bibinfo {volume} {273}},\ \bibinfo {pages} {L101} (\bibinfo {year} {1983})}\BibitemShut {NoStop}%
\bibitem [{\citenamefont {Chen}\ and\ \citenamefont {Palmadesso}(1984)}]{Chen1984}%
  \BibitemOpen
  \bibfield  {author} {\bibinfo {author} {\bibfnamefont {J.}~\bibnamefont {Chen}}\ and\ \bibinfo {author} {\bibfnamefont {P.}~\bibnamefont {Palmadesso}},\ }\bibfield  {title} {\bibinfo {title} {Tearing instability in an anisotropic neutral sheet},\ }\href {https://doi.org/10.1063/1.864727} {\bibfield  {journal} {\bibinfo  {journal} {Physics of Fluids}\ }\textbf {\bibinfo {volume} {27}},\ \bibinfo {pages} {1198} (\bibinfo {year} {1984})}\BibitemShut {NoStop}%
\bibitem [{\citenamefont {Karimabadi}\ \emph {et~al.}(2004)\citenamefont {Karimabadi}, \citenamefont {Daughton},\ and\ \citenamefont {Quest}}]{Karimabadi2004}%
  \BibitemOpen
  \bibfield  {author} {\bibinfo {author} {\bibfnamefont {H.}~\bibnamefont {Karimabadi}}, \bibinfo {author} {\bibfnamefont {W.}~\bibnamefont {Daughton}},\ and\ \bibinfo {author} {\bibfnamefont {K.~B.}\ \bibnamefont {Quest}},\ }\bibfield  {title} {\bibinfo {title} {Role of electron temperature anisotropy in the onset of magnetic reconnection},\ }\href {https://doi.org/10.1029/2004GL020791} {\bibfield  {journal} {\bibinfo  {journal} {Geophysical Research Letters}\ }\textbf {\bibinfo {volume} {31}},\ \bibinfo {pages} {L18801} (\bibinfo {year} {2004})}\BibitemShut {NoStop}%
\bibitem [{\citenamefont {Quest}\ \emph {et~al.}(2010)\citenamefont {Quest}, \citenamefont {Karimabadi},\ and\ \citenamefont {Daughton}}]{Quest2010}%
  \BibitemOpen
  \bibfield  {author} {\bibinfo {author} {\bibfnamefont {K.~B.}\ \bibnamefont {Quest}}, \bibinfo {author} {\bibfnamefont {H.}~\bibnamefont {Karimabadi}},\ and\ \bibinfo {author} {\bibfnamefont {W.}~\bibnamefont {Daughton}},\ }\bibfield  {title} {\bibinfo {title} {Linear theory of anisotropy driven modes in a {Harris} neutral sheet},\ }\href {https://doi.org/10.1063/1.3309731} {\bibfield  {journal} {\bibinfo  {journal} {Physics of Plasmas}\ }\textbf {\bibinfo {volume} {17}},\ \bibinfo {pages} {022107} (\bibinfo {year} {2010})}\BibitemShut {NoStop}%
\bibitem [{\citenamefont {Zenitani}\ and\ \citenamefont {Hesse}(2008)}]{Zenitani2008}%
  \BibitemOpen
  \bibfield  {author} {\bibinfo {author} {\bibfnamefont {S.}~\bibnamefont {Zenitani}}\ and\ \bibinfo {author} {\bibfnamefont {M.}~\bibnamefont {Hesse}},\ }\bibfield  {title} {\bibinfo {title} {The role of the {Weibel} instability at the reconnection jet front in relativistic pair plasma reconnection},\ }\href {https://doi.org/10.1063/1.2836623} {\bibfield  {journal} {\bibinfo  {journal} {Physics of Plasmas}\ }\textbf {\bibinfo {volume} {15}},\ \bibinfo {pages} {022101} (\bibinfo {year} {2008})}\BibitemShut {NoStop}%
\bibitem [{\citenamefont {Liu}\ \emph {et~al.}(2009)\citenamefont {Liu}, \citenamefont {Swisdak},\ and\ \citenamefont {Drake}}]{Liu2009}%
  \BibitemOpen
  \bibfield  {author} {\bibinfo {author} {\bibfnamefont {Y.-H.}\ \bibnamefont {Liu}}, \bibinfo {author} {\bibfnamefont {M.}~\bibnamefont {Swisdak}},\ and\ \bibinfo {author} {\bibfnamefont {J.~F.}\ \bibnamefont {Drake}},\ }\bibfield  {title} {\bibinfo {title} {The {Weibel} instability inside the electron-positron {Harris} sheet},\ }\href {https://doi.org/10.1063/1.3097474} {\bibfield  {journal} {\bibinfo  {journal} {Physics of Plasmas}\ }\textbf {\bibinfo {volume} {16}},\ \bibinfo {pages} {042101} (\bibinfo {year} {2009})}\BibitemShut {NoStop}%
\bibitem [{\citenamefont {Bulanov}\ \emph {et~al.}(1978)\citenamefont {Bulanov}, \citenamefont {Syrovatskii},\ and\ \citenamefont {Sakai}}]{Bulanov1978}%
  \BibitemOpen
  \bibfield  {author} {\bibinfo {author} {\bibfnamefont {S.~V.}\ \bibnamefont {Bulanov}}, \bibinfo {author} {\bibfnamefont {S.~I.}\ \bibnamefont {Syrovatskii}},\ and\ \bibinfo {author} {\bibfnamefont {J.}~\bibnamefont {Sakai}},\ }\bibfield  {title} {\bibinfo {title} {Stabilizing influence of plasma flow on dissipative tearing instability},\ }\href {https://ui.adsabs.harvard.edu/abs/1978JETPL..28..177B} {\bibfield  {journal} {\bibinfo  {journal} {JETP Letters}\ }\textbf {\bibinfo {volume} {28}},\ \bibinfo {pages} {177} (\bibinfo {year} {1978})}\BibitemShut {NoStop}%
\bibitem [{\citenamefont {Spitkovsky}(2005)}]{Spitkovsky2005}%
  \BibitemOpen
  \bibfield  {author} {\bibinfo {author} {\bibfnamefont {A.}~\bibnamefont {Spitkovsky}},\ }\bibfield  {title} {\bibinfo {title} {Simulations of relativistic collisionless shocks: shock structure and particle acceleration},\ }in\ \href {https://doi.org/10.1063/1.2141897} {\emph {\bibinfo {booktitle} {Astrophysical Sources of High Energy Particles and Radiation}}},\ \bibinfo {series} {AIP Conference Proceedings}, Vol.\ \bibinfo {volume} {801},\ \bibinfo {editor} {edited by\ \bibinfo {editor} {\bibfnamefont {T.}~\bibnamefont {Bulik}}, \bibinfo {editor} {\bibfnamefont {B.}~\bibnamefont {Rudak}},\ and\ \bibinfo {editor} {\bibfnamefont {G.}~\bibnamefont {Madejski}}}\ (\bibinfo {year} {2005})\ pp.\ \bibinfo {pages} {345--350}\BibitemShut {NoStop}%
\bibitem [{\citenamefont {Liu}\ \emph {et~al.}(2012)\citenamefont {Liu}, \citenamefont {Drake},\ and\ \citenamefont {Swisdak}}]{Liu2012}%
  \BibitemOpen
  \bibfield  {author} {\bibinfo {author} {\bibfnamefont {Y.-H.}\ \bibnamefont {Liu}}, \bibinfo {author} {\bibfnamefont {J.~F.}\ \bibnamefont {Drake}},\ and\ \bibinfo {author} {\bibfnamefont {M.}~\bibnamefont {Swisdak}},\ }\bibfield  {title} {\bibinfo {title} {The structure of the magnetic reconnection exhaust boundary},\ }\href {https://doi.org/10.1063/1.3685755} {\bibfield  {journal} {\bibinfo  {journal} {Physics of Plasmas}\ }\textbf {\bibinfo {volume} {19}},\ \bibinfo {pages} {022110} (\bibinfo {year} {2012})}\BibitemShut {NoStop}%
\bibitem [{\citenamefont {Higashimori}\ and\ \citenamefont {Hoshino}(2012)}]{Higashimori2012}%
  \BibitemOpen
  \bibfield  {author} {\bibinfo {author} {\bibfnamefont {K.}~\bibnamefont {Higashimori}}\ and\ \bibinfo {author} {\bibfnamefont {M.}~\bibnamefont {Hoshino}},\ }\bibfield  {title} {\bibinfo {title} {The relation between ion temperature anisotropy and formation of slow shocks in collisionless magnetic reconnection},\ }\href {https://doi.org/10.1029/2011JA016817} {\bibfield  {journal} {\bibinfo  {journal} {Journal of Geophysical Research: Space Physics}\ }\textbf {\bibinfo {volume} {117}},\ \bibinfo {pages} {A01220} (\bibinfo {year} {2012})}\BibitemShut {NoStop}%
\bibitem [{\citenamefont {Innocenti}\ \emph {et~al.}(2015)\citenamefont {Innocenti}, \citenamefont {Goldman}, \citenamefont {Newman}, \citenamefont {Markidis},\ and\ \citenamefont {Lapenta}}]{Innocenti2015}%
  \BibitemOpen
  \bibfield  {author} {\bibinfo {author} {\bibfnamefont {M.~E.}\ \bibnamefont {Innocenti}}, \bibinfo {author} {\bibfnamefont {M.}~\bibnamefont {Goldman}}, \bibinfo {author} {\bibfnamefont {D.}~\bibnamefont {Newman}}, \bibinfo {author} {\bibfnamefont {S.}~\bibnamefont {Markidis}},\ and\ \bibinfo {author} {\bibfnamefont {G.}~\bibnamefont {Lapenta}},\ }\bibfield  {title} {\bibinfo {title} {Evidence of magnetic field switch-off in collisionless magnetic reconnection},\ }\href {https://doi.org/10.1088/2041-8205/810/2/L19} {\bibfield  {journal} {\bibinfo  {journal} {The Astrophysical Journal Letters}\ }\textbf {\bibinfo {volume} {810}},\ \bibinfo {pages} {L19} (\bibinfo {year} {2015})}\BibitemShut {NoStop}%
\bibitem [{\citenamefont {Schoeffler}\ \emph {et~al.}(2026)\citenamefont {Schoeffler}, \citenamefont {Aravindakshan}, \citenamefont {Pucci},\ and\ \citenamefont {Innocenti}}]{Schoeffler2026}%
  \BibitemOpen
  \bibfield  {author} {\bibinfo {author} {\bibfnamefont {K.~M.}\ \bibnamefont {Schoeffler}}, \bibinfo {author} {\bibfnamefont {H.}~\bibnamefont {Aravindakshan}}, \bibinfo {author} {\bibfnamefont {F.}~\bibnamefont {Pucci}},\ and\ \bibinfo {author} {\bibfnamefont {M.~E.}\ \bibnamefont {Innocenti}},\ }\href@noop {} {\bibinfo {title} {Are the tearing and the {Weibel} instabilities the same?}} (\bibinfo {year} {2026}),\ \Eprint {https://arxiv.org/abs/2611.09639} {arXiv:2611.09639 [physics.plasm-ph]} \BibitemShut {NoStop}%
\bibitem [{\citenamefont {Karimabadi}\ \emph {et~al.}(2005)\citenamefont {Karimabadi}, \citenamefont {Daughton},\ and\ \citenamefont {Quest}}]{Karimabadi2005}%
  \BibitemOpen
  \bibfield  {author} {\bibinfo {author} {\bibfnamefont {H.}~\bibnamefont {Karimabadi}}, \bibinfo {author} {\bibfnamefont {W.}~\bibnamefont {Daughton}},\ and\ \bibinfo {author} {\bibfnamefont {K.~B.}\ \bibnamefont {Quest}},\ }\bibfield  {title} {\bibinfo {title} {Physics of saturation of collisionless tearing mode as a function of guide field},\ }\href {https://doi.org/10.1029/2004JA010749} {\bibfield  {journal} {\bibinfo  {journal} {Journal of Geophysical Research: Space Physics}\ }\textbf {\bibinfo {volume} {110}},\ \bibinfo {pages} {A03214} (\bibinfo {year} {2005})}\BibitemShut {NoStop}%
\bibitem [{\citenamefont {Daughton}\ \emph {et~al.}(2011)\citenamefont {Daughton}, \citenamefont {Roytershteyn}, \citenamefont {Karimabadi}, \citenamefont {Yin}, \citenamefont {Albright}, \citenamefont {Bergen},\ and\ \citenamefont {Bowers}}]{Daughton2011}%
  \BibitemOpen
  \bibfield  {author} {\bibinfo {author} {\bibfnamefont {W.}~\bibnamefont {Daughton}}, \bibinfo {author} {\bibfnamefont {V.}~\bibnamefont {Roytershteyn}}, \bibinfo {author} {\bibfnamefont {H.}~\bibnamefont {Karimabadi}}, \bibinfo {author} {\bibfnamefont {L.}~\bibnamefont {Yin}}, \bibinfo {author} {\bibfnamefont {B.~J.}\ \bibnamefont {Albright}}, \bibinfo {author} {\bibfnamefont {B.}~\bibnamefont {Bergen}},\ and\ \bibinfo {author} {\bibfnamefont {K.~J.}\ \bibnamefont {Bowers}},\ }\bibfield  {title} {\bibinfo {title} {Role of electron physics in the development of turbulent magnetic reconnection in collisionless plasmas},\ }\href {https://doi.org/10.1038/nphys1965} {\bibfield  {journal} {\bibinfo  {journal} {Nature Physics}\ }\textbf {\bibinfo {volume} {7}},\ \bibinfo {pages} {539} (\bibinfo {year} {2011})}\BibitemShut {NoStop}%
\bibitem [{\citenamefont {Sironi}\ and\ \citenamefont {Beloborodov}(2020)}]{Sironi2020}%
  \BibitemOpen
  \bibfield  {author} {\bibinfo {author} {\bibfnamefont {L.}~\bibnamefont {Sironi}}\ and\ \bibinfo {author} {\bibfnamefont {A.~M.}\ \bibnamefont {Beloborodov}},\ }\bibfield  {title} {\bibinfo {title} {Kinetic simulations of radiative magnetic reconnection in the coronae of accreting black holes},\ }\href {https://doi.org/10.3847/1538-4357/aba622} {\bibfield  {journal} {\bibinfo  {journal} {The Astrophysical Journal}\ }\textbf {\bibinfo {volume} {899}},\ \bibinfo {pages} {52} (\bibinfo {year} {2020})}\BibitemShut {NoStop}%
\bibitem [{\citenamefont {Moran}\ \emph {et~al.}(2025)\citenamefont {Moran}, \citenamefont {Sironi}, \citenamefont {Levis}, \citenamefont {Ripperda}, \citenamefont {Most},\ and\ \citenamefont {Selvi}}]{Moran2025}%
  \BibitemOpen
  \bibfield  {author} {\bibinfo {author} {\bibfnamefont {A.}~\bibnamefont {Moran}}, \bibinfo {author} {\bibfnamefont {L.}~\bibnamefont {Sironi}}, \bibinfo {author} {\bibfnamefont {A.}~\bibnamefont {Levis}}, \bibinfo {author} {\bibfnamefont {B.}~\bibnamefont {Ripperda}}, \bibinfo {author} {\bibfnamefont {E.~R.}\ \bibnamefont {Most}},\ and\ \bibinfo {author} {\bibfnamefont {S.}~\bibnamefont {Selvi}},\ }\bibfield  {title} {\bibinfo {title} {Effective resistivity in relativistic reconnection: A prescription based on fully kinetic simulations},\ }\href {https://doi.org/10.3847/2041-8213/ada158} {\bibfield  {journal} {\bibinfo  {journal} {The Astrophysical Journal Letters}\ }\textbf {\bibinfo {volume} {978}},\ \bibinfo {pages} {L45} (\bibinfo {year} {2025})}\BibitemShut {NoStop}%
\bibitem [{\citenamefont {Ripperda}\ \emph {et~al.}(2026)\citenamefont {Ripperda}, \citenamefont {Grehan}, \citenamefont {Moran}, \citenamefont {Selvi}, \citenamefont {Sironi}, \citenamefont {Philippov}, \citenamefont {Bransgrove},\ and\ \citenamefont {Porth}}]{Ripperda2026}%
  \BibitemOpen
  \bibfield  {author} {\bibinfo {author} {\bibfnamefont {B.}~\bibnamefont {Ripperda}}, \bibinfo {author} {\bibfnamefont {M.~P.}\ \bibnamefont {Grehan}}, \bibinfo {author} {\bibfnamefont {A.}~\bibnamefont {Moran}}, \bibinfo {author} {\bibfnamefont {S.}~\bibnamefont {Selvi}}, \bibinfo {author} {\bibfnamefont {L.}~\bibnamefont {Sironi}}, \bibinfo {author} {\bibfnamefont {A.}~\bibnamefont {Philippov}}, \bibinfo {author} {\bibfnamefont {A.}~\bibnamefont {Bransgrove}},\ and\ \bibinfo {author} {\bibfnamefont {O.}~\bibnamefont {Porth}},\ }\href@noop {} {\bibinfo {title} {Magnetic reconnection with a 0.1 rate: Effective resistivity in general relativistic magnetohydrodynamics}} (\bibinfo {year} {2026}),\ \Eprint {https://arxiv.org/abs/2601.02460} {arXiv:2601.02460 [astro-ph.HE]} \BibitemShut {NoStop}%
\end{thebibliography}%

\appendix
\section{Appendix A: Derivation of the Residence-Time Distribution}
\label{app:residence}

Near a flow-dividing stagnation point ($x = 0$), the lowest-order odd outflow
velocity profile across the diffusion region $|x| \le L_x$ is linear,
\begin{equation}
v_x(x) = \frac{v_{\rm out}}{L_x}\,x \equiv \frac{x}{\tau_{\rm res}},
\label{eq:vx_linear}
\end{equation}
with constant divergence $\partial_x v_x = \tau_{\rm res}^{-1}$, fed by a
spatially uniform inflow of newly arriving ($\tau = 0$) particles from $\pm y$
at volumetric rate $Q = n_0 v_{\rm in}/\delta$ (for upstream thermal speed
$v_{x0} \ll v_{\rm out}$). A particle entering the unmagnetized meandering
layer at initial position $x_0$ at age $\tau = 0$ obeys
$dx/d\tau = x/\tau_{\rm res}$, yielding the deterministic trajectory
\begin{equation}
x(\tau) = x_0\, e^{\tau/\tau_{\rm res}}.
\label{eq:x_traj}
\end{equation}

This trajectory stretching gives the same exponential distribution for both the
total exit time and the local age at any position $x$, shown below. 

\emph{Exit (residence-time) distribution.---}A particle entering at
$x_0 \in (0, L_x]$ reaches the boundary $x = L_x$ after total residence time
$\tau_{\rm exit}(x_0) = \tau_{\rm res}\ln(L_x/x_0)$, or equivalently
$x_0(\tau_{\rm exit}) = L_x e^{-\tau_{\rm exit}/\tau_{\rm res}}$ (by symmetry,
the same holds for $x_0 < 0$). Because inflow is uniform in $x_0$, the fraction
of entering particles that remain in the diffusion region longer than $\tau$ is
\begin{equation}
\Pr(\tau_{\rm exit} > \tau) = \frac{x_0(\tau)}{L_x} = e^{-\tau/\tau_{\rm res}}.
\label{eq:survival}
\end{equation}
Differentiating yields the normalized residence-time probability density
\begin{equation}
P(\tau) = -\frac{d}{d\tau}\!\left(e^{-\tau/\tau_{\rm res}}\right) = \frac{1}{\tau_{\rm res}} e^{-\tau/\tau_{\rm res}} \qquad (\tau \ge 0).
\label{eq:P_tau}
\end{equation}
Because $L_x$ cancels in Eq.~(\ref{eq:survival}), the identical distribution
applies to any narrower strip $|x| \le w$ centered on $x = 0$.

\emph{Local age distribution.---}Particles passing through $[x, x+dx]$ with
ages in $[\tau, \tau+d\tau]$ must have entered at
$x_0 = x e^{-\tau/\tau_{\rm res}}$ within an initial interval
$dx_0 = e^{-\tau/\tau_{\rm res}}\,dx$. Thus the diverging outflow stretches
each fluid element by $e^{+\tau/\tau_{\rm res}}$ and dilutes its density by
$e^{-\tau/\tau_{\rm res}}$, giving a local density of particles of age $\tau$,
\begin{equation}
n(x,\tau)\,d\tau = \frac{Q\,dx_0\,d\tau}{dx} = Q\,e^{-\tau/\tau_{\rm res}}\,d\tau,
\label{eq:n_local}
\end{equation}
independent of $x$. Normalizing by the total local density
$n_{\rm tot} = \int_0^\infty Q e^{-\tau/\tau_{\rm res}}\,d\tau = Q\tau_{\rm res}$
recovers $P(\tau) = \tau_{\rm res}^{-1} e^{-\tau/\tau_{\rm res}}$ at every
position inside the outflow region.

\section{Appendix B: Derivation of the Out-of-Plane Momentum Distribution}
\label{app:laplace}

In a neutral pair plasma of total density $n_0 = n_+ + n_-$ at the X-point,
positrons ($+$) and electrons ($-$) each have density $n_\pm = n_0/2$ and local
age distribution $P(\tau) = \tau_{\rm res}^{-1} e^{-\tau/\tau_{\rm res}}$
(\hyperref[app:residence]{Appendix~A}). The number density of each species with
ages in $[\tau, \tau + d\tau]$ ($d\tau > 0$) is therefore
$dn_\pm = (n_0/2) P(\tau)\,d\tau$. Neglecting the small upstream thermal spread
($|p_{z0}| \ll p_d$), a particle of charge $q = \pm e$ entering with
$p_z(0) \approx 0$ accelerates linearly in $E_z$ (taking $E_z > 0$; reversing
the sign simply swaps the two species):
\begin{equation}
p_z(\tau) = q E_z \tau = \pm p_d\,\frac{\tau}{\tau_{\rm res}}, \qquad p_d \equiv e E_z \tau_{\rm res}.
\label{eq:pz_tau}
\end{equation}

This monotonic relation maps $\tau \in [0, \infty)$ one-to-one onto disjoint
half-lines in momentum space, with $\tau = \tau_{\rm res}|p_z|/p_d$ and
positive bin width $|dp_z| = (p_d/\tau_{\rm res})\,d\tau$. Equating particle
counts in corresponding bins, $f_\pm(p_z)\,|dp_z| = n_\pm P(\tau)\,d\tau$,
gives
\begin{equation}
f_\pm(p_z) = n_\pm P(\tau)\left|\frac{d\tau}{dp_z}\right|
= \frac{n_0}{2 p_d} e^{\mp p_z/p_d}\,H(\pm p_z),
\label{eq:f_pm_deriv}
\end{equation}
where $H$ is the Heaviside step function ($f_+ = 0$ for $p_z < 0$ and
$f_- = 0$ for $p_z > 0$, because ages satisfy $\tau \ge 0$). Since at any
$p_z \neq 0$ only one species contributes, summing both species yields the
continuous Laplace distribution of Eq.~(\ref{eq:laplace_dist}),
$F(p_z) \equiv f_+(p_z) + f_-(p_z) = (n_0/2p_d) e^{-|p_z|/p_d}$, which
normalizes to $\int_{-\infty}^\infty F(p_z)\,dp_z = n_0$ and has moments
$\langle p_z\rangle_\pm = \pm p_d$ and $\langle p_z^2\rangle = 2 p_d^2$.

\section{Appendix C: Effective Electric Field and Moments of the Momentum Cap}
\label{app:cap}

Because the diverging outflow sweeps particles away from $x = 0$, any particle
inside the narrow capping strip $|x| \le w$ has spent its entire prior residence
time $\tau$ within the strip and thus retains the exponential age distribution
$P(\tau) = \tau_{\rm res}^{-1} e^{-\tau/\tau_{\rm res}}$ derived in
\hyperref[app:residence]{Appendix~A}. Under the momentum ceiling
$|p_z| \le p_{\rm cap} \equiv \xi p_d$, particles in the bulk current-carrying
core ($\tau < \xi\tau_{\rm res}$, fraction $1 - e^{-\xi}$) accelerate freely as
$p_z(\tau) = q E_z \tau$ and experience no clamping force ($F_{\rm cap} = 0$).
Only the oldest tail particles ($\tau \ge \xi\tau_{\rm res}$, fraction
$\int_{\xi\tau_{\rm res}}^\infty P(\tau)\,d\tau = e^{-\xi}$) reach the ceiling
and remain pinned at $|p_z| = \xi p_d$ ($dp_z/dt = 0$), requiring a counter-force
$F_{\rm cap} = -q E_z$ to cancel their electric acceleration.

Setting $s \equiv \tau/\tau_{\rm res}$ and averaging the effective electric
field $E_{\rm cap} \equiv -\langle F_{\rm cap}/q\rangle$ and capped profile
$|p_z| = p_d \min(s, \xi)$ over $P(\tau)$ gives
\begin{equation}
\begin{aligned}
\frac{E_{\rm cap}}{E_z} &= \int_\xi^\infty e^{-s}\,ds = e^{-\xi}, \\[2pt]
\frac{\langle |p_z|\rangle}{p_d} &= \int_0^\xi s e^{-s}\,ds + \xi e^{-\xi} = 1 - e^{-\xi}, \\[2pt]
\frac{\langle p_z^2\rangle}{2 p_d^2} &= \frac{1}{2}\!\left(\int_0^\xi s^2 e^{-s}\,ds + \xi^2 e^{-\xi}\right) \\
&= 1 - (1+\xi)e^{-\xi}.
\end{aligned}
\label{eq:cap_moments}
\end{equation}
For our fiducial cap $\xi = 2.3$ ($p_{\rm cap} = 2.3 p_d$),
Eq.~(\ref{eq:cap_moments}) gives $E_{\rm cap}/E_z = e^{-2.3} \approx 0.10$,
$\langle |p_z|\rangle/p_d \approx 0.90$, and
$\langle p_z^2\rangle/(2 p_d^2) = 1 - 3.3 e^{-2.3} \approx 0.67$, a $33\%$
reduction in $\langle p_z^2\rangle$ at only $10\%$ cost to the bulk current
moment.

\end{document}